# An incremental algorithm based on multichannel non-negative matrix partial co-factorization for ambient denoising in auscultation

Juan De La Torre Cruz *, Francisco Jesús Cañadas Quesada, Damián Martínez-Muñoz, Nicolás Ruiz Reyes, Sebastián García Galán, Julio José Carabias Orti

*Departament of Telecommunication Engineering. University of Jaen, Campus Cientifico-Tecnologico de Linares, Avda. de la Universidad, s/n, 23700 Linares, Jaen, Spain*

**Abstract**

One of the major current limitations in the diagnosis derived from auscultation remains the ambient noise surrounding the subject, which prevents successful auscultation. Therefore, it is essential to develop robust signal processing algorithms that can extract relevant clinical information from auscultated recordings analyzing in depth the acoustic environment in order to help the decision-making process made by physicians. The aim of this study is to implement a method to remove ambient noise in biomedical sounds captured in auscultation. We propose an incremental approach based on multichannel non-negative matrix partial co-factorization (NMPCF) for ambient denoising focusing on high noisy environment with a Signal-to-Noise Ratio (SNR) ≤ -5 dB. The first contribution applies NMPCF assuming that ambient noise can be modelled as repetitive sound events simultaneously found in two single-channel inputs captured by means of different recording devices. The second contribution proposes an incremental algorithm, based on the previous multichannel NMPCF, that refines the estimated biomedical spectrogram throughout a set of incremental stages

*Corresponding author. Tlf.: (+34) 953648592

*Email addresses:* jtorre@ujaen.es (Juan De La Torre Cruz *), fcanadas@ujaen.es (Francisco Jesu´s Cañadas Quesada), damian@ujaen.es (Damián Martínez-Muñoz), nicolas@ujaen.es (Nicolás Ruiz Reyes), sgalan@ujaen.es (Sebastián García Galán), carabias@ujaen.es (Julio José Carabias Orti)

by eliminating most of the ambient noise that was not removed in the previous stage at the expense of preserving most of the biomedical spectral content. The ambient denoising performance of the proposed method, compared to some of the most relevant state-of-the-art methods, has been evaluated using a set of recordings composed of biomedical sounds mixed with ambient noise that typically surrounds a medical consultation room to simulate high noisy environments with a SNR from -20 dB to -5 dB. In order to analyse the drop in denoising performance of the evaluated methods when the effect of the propagation of the patient's body material and the acoustics of the room is considered, results have been obtained with and without taking these effects into account. Experimental results report that: (i) the performance drop suffered by the proposed method is lower compared to MSS and NLMS when considering the effect of the propagation of the patient's body material and the acoustics of the room active; (ii) unlike what happens with MSS and NLMS, the proposed method shows a stable trend of the average SDR and SIR results regardless of the type of ambient noise and the SNR level evaluated; and (iii) a remarkable advantage of the proposed method is the high robustness of the acoustic quality of the estimated biomedical sounds when the two single-channel inputs suffer from a delay between them.

## 1. Introduction

Auscultation is defined as the technique of listening the internal sounds produced by the human organs by means of a stethoscope. This technique is simple, non-invasive, safe and inexpensive that provides valuable clinical information in the diagnosis of the status of the heart, lung and airways [1, 2]. Although today there are more advanced medical tools to analyze the status of the heart and lung, such as chest radiography, electrocardiography (ECG), spirometry or laboratory analyses, auscultation is still one of the most widely used techniques to detect any cardiac or pulmonary disease. However, the diagnosis derived from auscultation shows two main limitations: i) high subjectivity due to the physician's expertise to recognize sounds that reveal any physiological disorder [3]; ii) high dependence on the ambient noise surrounding the subject to provide a reliable diagnosis [4].

Because the process of auscultation in a soundproof room is not possible in most cases, especially in low-income and middle-income countries supported by a resource-poor health system, ambient denoising performed in the examination room of the health center is still a challenging task in biomedical signal processing in order to maximize the reliability of a diagnosis. The main effects caused by ambient noise are the masking, distortion and weakness of the sound of interest that may provide relevant clues in the diagnosis as shown in Figure 1. As a result, the probability of making a medical error increases when auscultation is performed in a noisy environment since the physician is not able to correctly interpret the diagnostic information contained in the sound signal from auscultation. In this work, the term biomedical means that the sound sources that have generated the sounds of interest have been the human internal organs, and specifically, the heart and the lung.

In recent years, several signal processing tasks have been applied in the field of biomedical information retrieval such as, sound source separation [5, 6, 7] as well as sound event detection [8, 9, 10, 11] and classification [12, 13, 14, 15, 16]. However, most of their experimental results have been obtained in environments in which the biomedical sounds are not acoustically contaminated by ambient noises. Therefore, the task of ambient denoising is still an open research topic in biomedical engineering being most of the approaches based on adaptive filtering [17, 18, 19, 20, 21] and spectral subtraction [22, 23, 24, 25]. Chang and Lai [23] proposed a two-channel spectral subtraction method, based on autoregressive (AR), mel-frequency cepstral coefficients (MFCC) and dynamic time warping (DTW), applied to the lung sound signals under noisy conditions before the extraction of lung sound features. Emmanouilidou et al. [24] developed a multiband spectral scheme, based on two-microphone setup, to suppress the background noise while successfully preserving the lung sound content to maximize the informative diagnostic value obtained from auscultation. The algorithm analyzes each frequency band in a nonuniform manner and uses prior knowledge of the target sounds to apply a penalty in the spectral domain. It follows from the above that it is crucial to develop robust signal processing algorithms that can extract relevant clinical information from auscultated recordings taking into consideration the acoustic environment that surrounds the subject in order to improve the decision making process made by physicians.

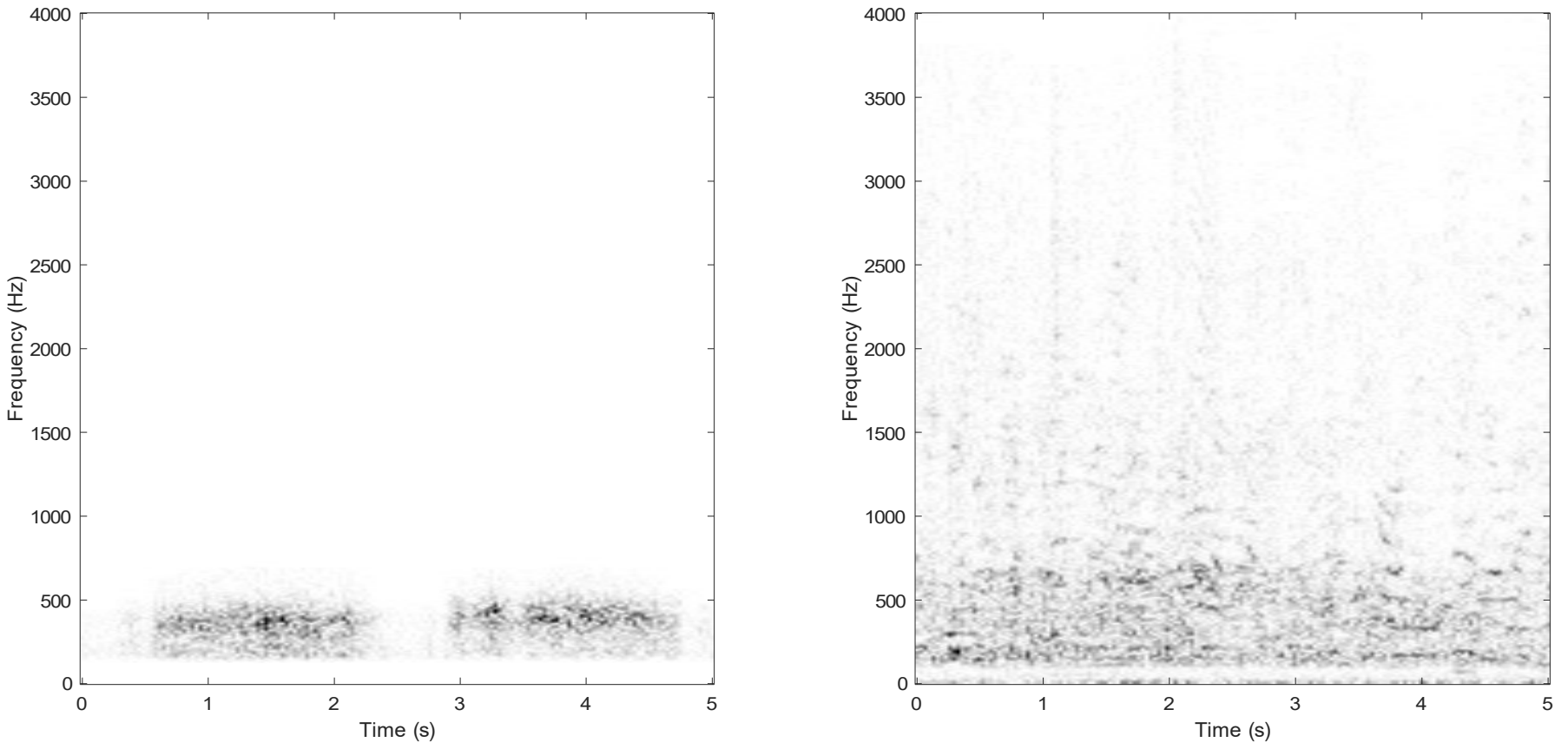

Figure 1: Spectrogram from a clean lung sound recording: (Left) under no ambient noise; (Right) mixed with ambient noise (babble) in a Signal-to-Noise Ratio (SNR) equals -10 dB. Comparing both figures, it can be observed that the spectral content from lung is completely masked by ambient noise. Higher energies are indicated by darker colour.

It is well known that the conventional Non-negative Matrix Partial Co-Factorization (NMPCF) enforces a joint matrix decomposition using multiple matrices to recover a set of shared spectral patterns (bases) that model the spectral behavior of some of the sound sources contained in the single-channel input. Over the last decade, NMPCF has been successfully applied in several single-channel sound signal processing fields: i) Music Information Retrieval (MIR) such as, singing-voice separation [26], rhythmic extraction [27, 28, 29] and speaker diarization [30]; and ii) Enhancement of biomedical sounds such as, normal respiratory and wheezes sound separation [31]. In this work, we pro-pose an incremental algorithm, called 2C-NMPCF, that improves the quality of biomedical sounds captured in auscultated recordings by applying the conventional NMPCF from a multi-channel scenario rather than a single-channel. In this paper, the term multichannel refers to the use of two single-channel audio inputs simultaneously captured by means of different recording devices. As occurs in [24], these two single-channel inputs are defined as: (i) the internal recording that comes from the audio captured using a stethoscope in which both biomedical sounds from inside the human body and ambient noises can be listened; (ii) the external recording that comes from the audio captured using an external microphone in which only the ambient noise that surrounds the subject is captured. Specifically, our first contribution applies NMPCF from a multichannel point of view assuming that ambient noises can be modelled as repetitive sounds that can be simultaneously found in both single-channel inputs. In other words, we implicitly assume that the spectral patterns that characterize the ambient noises are repeated sound events contained in both the spectrograms from the internal and external recordings. Our second contribution proposes an incremental algorithm, based on the previous multichannel NMPCF, that refines the estimated biomedical spectrogram through a set of incremental stages by eliminating a high amount of ambient noise that was not extracted in the previous stage, especially in the case of high noise environments. In this work, a high noisy environment provides a Signal-to-Noise Ratio (SNR) lower than 0 dB.

The paper is structured as follows: Section 2 details the datasets, the state-of-the-art methods for comparison and the proposed method. The metrics, setup and results are shown in Section 3. Finally, Section 4 presents the conclusions and future work.

## 2. Materials and Methods

### 2.1. Data collection

Due to the lack of publicly available databases consisting of biomedical sounds mixed with ambient noises to the best of our knowledge, we have created the database $D_C$. The database $D_C$ is composed by the ambient noise database $D_N$ and the biomedical database $D_B$ in order to simulate auscultation recordings captured from a stethoscope.

The database $D_N$ has been created taking into account a wide range of ambient noises collected from databases widely used in the field of sound source separation [32] and sound event detection [33, 34]. Most of these ambient noises have been classified as some of the most disturbing noises that can appear in the auscultation performed in the hospital room according to information provided by medical personnel from the Hospital of Jaen (Spain). For this reason, the database $D_N$ is composed of five types of ambient noise in order to assess the denoising performance of the proposed method considering common indoor and outdoor ambient noises that typically surround a medical consultation room: ambulance siren [35, 36], baby crying [37], babble (people speaking) [38, 39], car (inside the vehicle) [40] and street (car passing by, car engine running, car idling, bus, truck, children yelling, people talking, workers on the street) [41, 42]. The database $D_N$ consists of a total of 150 single-channel recordings of ambient noises, of which each type of noise consists of 30 recordings. Each recording has a duration of 5 seconds that has been obtained applying a pseudo-random process, based on the standard uniform distribution, to select a starting time followed by a 5 seconds interval.

The database $D_B$ consists of a total of 150 single-channel biomedical recordings from public and private biomedical databases, specifically, 75 heart recordings [43, 44] (typically in the range 10Hz-320 Hz [45, 46]) and 75 lung recordings [47] (typically in the range 50Hz-2500 Hz [48, 49]). Highlight that ambient noises are not listened on each recording. Each recording has a duration of 5 seconds which has been obtained applying a pseudo-random process similarly as used in the database $D_N$.

Each mixture recording belonging to the database $D_C$ has been generated mixing each recording from the database $D_B$ with a recording of each type of noise randomly chosen from the database $D_N$. Indicate that the recordings of noise used for the mixtures with the heart recordings are the same as those used for the mixing with the lung recordings. For each mixture recording from $D_C$, the ambient noise used in the internal recording is the same noise used in the external recording. The database $D_C$ is not affected by the effect of the patient's body material or by the acoustics room in order to perform a fair optimization parameters of the proposed method avoiding body-specific or room-specific optimization. As a result, two databases are created from $D_C$, the optimization database $D_O$ and the testing database $D_T$,

- The optimization database $D_O$ is generated randomly selecting two-thirds of all mixtures recordings from the database $D_C$.
- The testing database $D_T$ is generated using the remainder one-third mixtures recordings that are not used in the database $D_O$.

The set of recordings used in the optimization database $D_O$ is not the same as that used in the testing database $D_T$ in order to validate the denoising results. Moreover, several SNR have been applied in the mixing process to create the database $D_C$ in order to evaluate high noisy environments. In this way, the databases $D_{T_{-20}}$ (SNR=-20 dB), $D_{T_{-15}}$ (SNR=-15 dB), $D_{T_{-10}}$ (SNR=-10 dB) and $D_{T_{-5}}$ (SNR=-5 dB) refer to the same database $D_T$ but using different SNR between biomedical and ambient noise recordings. For example, a value SNR=-20 dB indicates that the power of the ambient noise is 100 times greater compared to the power of the biomedical sounds used in the audio mixture. Table 1 describes the characteristics of the data, according to the databases used in the experimental evaluation.

| $ID_1$ | $ID_2$ | $ID_3$ | $ID_4$ | $ID_5$ | $ID_6$ | $ID_7$ | $ID_8$ | $ID_9$ | $ID_{10}$ | $ID_{11}$ |
|---|---|---|---|---|---|---|---|---|---|---|
| $D_B$ | 75 | 75 | - | - | - | - | - | 150 | - | 750 |
| $D_N$ | - | - | 30 | 30 | 30 | 30 | 30 | - | 150 | 750 |
| $D_O$ | 50 | 50 | 20 | 20 | 20 | 20 | 20 | 100 | 100 | 2500 |
| $D_T$ | 25 | 25 | 10 | 10 | 10 | 10 | 10 | 50 | 50 | 1250 |

Table 1: Characteristics of the data. $ID_1$: database identifier; $ID_2$: number of clean heart recordings; $ID_3$: number of clean lung recordings; $ID_4$: number of ambulance siren noise recordings; $ID_5$: number of babycrying noise recordings; $ID_6$: number of babble noise recordings; $ID_7$: number of car noise recordings; $ID_8$: number of street noise recordings; $ID_9$: total number of biomedical (clean heart and lung) recordings; $ID_{10}$: total number of ambient noise recordings; $ID_{11}$: temporal duration for all recordings in seconds.

### *2.2. Baseline method for comparison*

A reference adaptive filtering method, Normalized Least-Mean-Square (NLMS) [50, 51], has been used to assess the performance of the proposed method in the task of ambient denoising. The optimal length of the filter has been optimized by varying the number of coefficients between 10 and 100 to maximize the denoising performance of the NLMS method. The optimization results showed that NLMS obtained the best denoising performance when the size of the adaptive filter is equal to 10. Moreover, one of the most relevant state-of-the-art methods, based on Multiband Spectral Subtraction (MSS) [24], has also been implemented. Note that the best configuration (algorithm B) of MSS has been chosen to perform a fair comparison. The reader can refer to [24] for more details.

### *2.3. Proposed method for ambient denoising*

The main problem that physicians point out when performing the auscultation process in high noisy environments is that the biomedical sounds are severely overlapped with ambient noises so, part of the valuable clinical information contained in the sounds of interest is masked. The aim of the proposed method is to improve the quality of the biomedical sounds captured by a stethoscope in high noisy environments applying Non-negative Matrix Partial

Co-Factorization (NMPCF) in a multichannel (two distinct single-channel signals) scenario (2C-NMPCF). The flowchart of the proposed method 2C-NMPCF is shown in Figure 2.

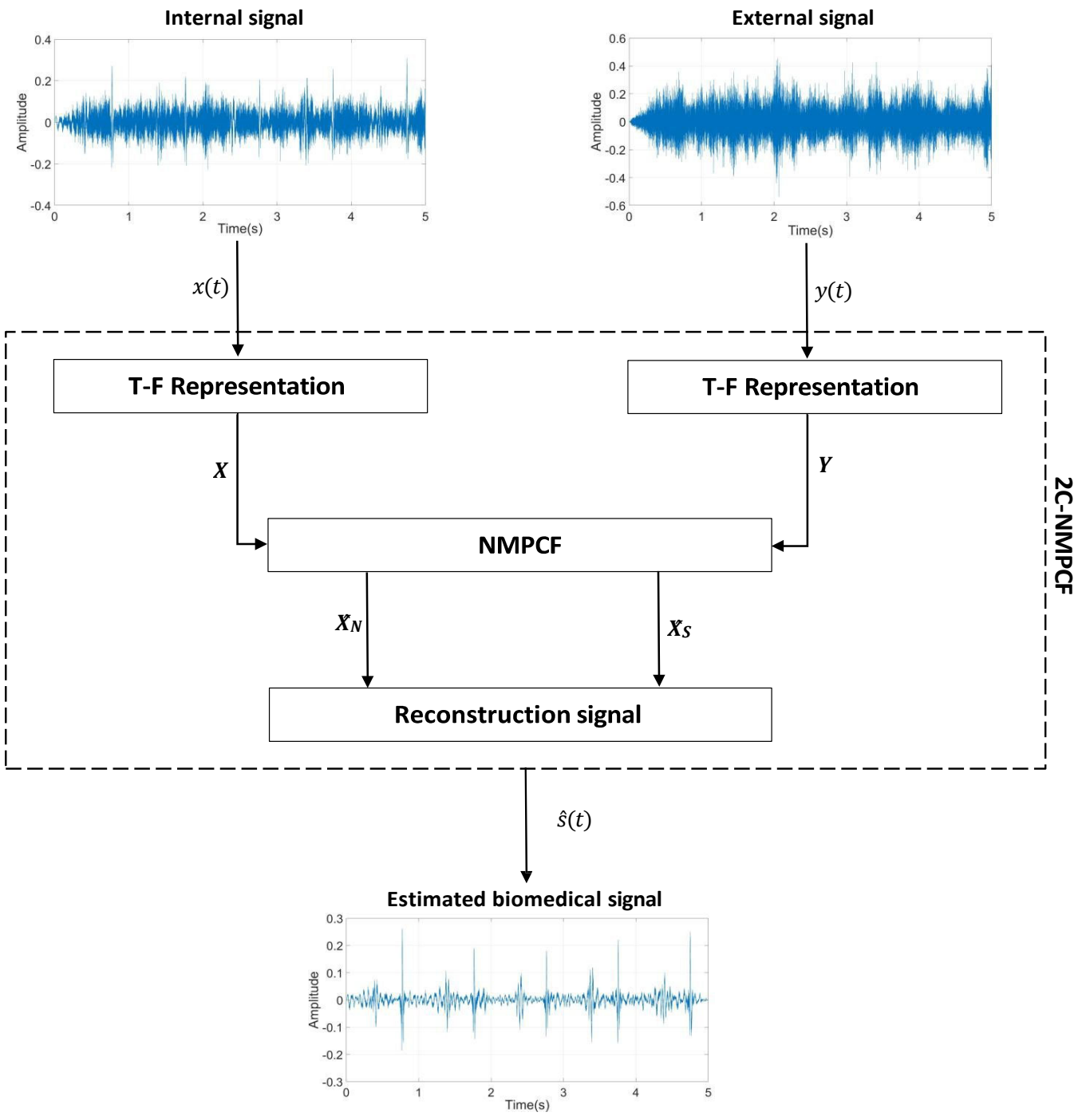

Figure 2: Flowchart of the proposed method 2C-NMPCF.

### *2.3.1. Time-Frequency representation*

The internal signal (first single-channel) $x(t)$ represents the sounds captured by a digital stethoscope that is composed of two types of additive sound sources: (i) the biomedical sounds $s(t)$ from the subject; (ii) the ambient noises $n(t)$ surrounding the subject that are still heard inside the human body. We assume that $s(t)$ and $n(t)$ can be considered independent sound sources, that is, $x(t) = s(t) + n(t)$. Moreover, an external microphone, located outside of the subject, captures all ambient noises that are represented by the external signal (second single-channel) $y(t)$.

The complex and magnitude spectrogram $\mathbf{X}_c \in \mathbb{C}_+^{F\times T}$, $\mathbf{X} \in \mathbb{R}_+^{F\times T}$ associated to the internal signal $x(t)$ and the magnitude spectrogram $\mathbf{Y} \in \mathbb{R}_+^{F\times T}$ associated to the external signal $y(t)$ are calculated using the Short-Time Fourier Transform (STFT) applying a Hamming window of size N with 50% overlap. Indicate that the complex values associated with $\mathbf{X}_c$, in which the phase information is included, are used later in the resynthesis process. The size and scale of the magnitude spectrograms depend on each input single-channel signal. Therefore, a normalization process is applied in order to ensure that the proposed method is independent of the size and scale of the input spectrograms. Thus, the normalized spectrograms $\overline{\mathbf{X}} \in \mathbb{R}_+^{F\times T}$ and $\overline{\mathbf{Y}} \in \mathbb{R}_+^{F\times T}$ are computed as follows,

$$\overline{\mathbf{Z}} = \frac{\mathbf{Z}}{\left(\frac{\sum_{f,t} Z_{f,t}}{FT}\right)} \tag{1}$$

where $\mathbf{Z} = \{\mathbf{X},\mathbf{Y}\}$ according to the input spectrogram. The variables $F$ and $T$ represent the number of frequency bins and the number of time frames. To avoid the complex nomenclature throughout the manuscript, the variables $\overline{\mathbf{X}}$ and $\overline{\mathbf{Y}}$ are hereinafter referred as $\mathbf{X}$ and $\mathbf{Y}$, respectively.

### *2.3.2. Multichannel Non-negative Matrix Partial Co-Factorization (2C-NMPCF)*

The idea of the proposed method 2C-NMPCF is to enforce a joint matrix decomposition using multiple matrices $\mathbf{X}$, $\mathbf{Y}$ obtained from distinct single-channel spectrograms instead of the several excerpts of the same single-channel spectrogram as occurs in the conventional NMPCF. The main contribution of the proposed method 2C-NMPCF is to exploit the spectral patterns that are shared in two distinct spectrograms since we assume that ambient noises can be modelled as repetitive sound events that can be simultaneously found in the spectrograms associated both the internal and external signal. This modeling allows to remove most of the ambient noises that are active in the internal signal improving the quality of the biomedical sounds from the auscultation process. The proposed method 2C-NMPCF is composed of two stages:

- Stage 1. This stage is applied to the internal signal $x(t)$. The input

spectrogram $\mathbf{X}$ is decomposed into two separated or estimated spectrograms, the magnitude spectrogram only composed of biomedical sounds $\hat{\mathbf{X}}_S \in \mathbb{R}_+^{F\times T}$ and the magnitude spectrogram only composed of ambient noises $\hat{\mathbf{X}}_N \in \mathbb{R}_+^{F\times T}$. The factorization of each spectrogram depends on the estimated basis matrix $\mathbf{U} \in \mathbb{R}_+^{F\times K}$ (dictionary of spectral patterns) and the estimated activation matrix $\mathbf{V} \in \mathbb{R}_+^{K\times T}$ (temporal gains) as follows,

$$\mathbf{X} \approx \hat{\mathbf{X}} = \hat{\mathbf{X}}_N + \hat{\mathbf{X}}_S = \mathbf{U}\mathbf{V} = \begin{bmatrix}\mathbf{U}_N & \mathbf{U}_S\end{bmatrix}\begin{bmatrix}\mathbf{V}_N \\ \mathbf{V}_S\end{bmatrix} = \mathbf{U}_N\mathbf{V}_N + \mathbf{U}_S\mathbf{V}_S \quad (2)$$

where $\hat{\mathbf{X}} \in \mathbb{R}_+^{F\times T}$ is the estimated or reconstructed magnitude spectrogram of the first channel signal. $\mathbf{U}_N \in \mathbb{R}_+^{F\times K_N}$ and $\mathbf{V}_N \in \mathbb{R}_+^{K_N\times T}$ are the estimated basis and activations matrix of the ambient noises. The variables $\mathbf{U}_S \in \mathbb{R}_+^{F\times K_S}$ and $\mathbf{V}_S \in \mathbb{R}_+^{K_S\times T}$ are the estimated basis and activation matrix of the biomedical sounds. The parameter $K = K_N + K_S$ indicates the number of bases, being $K_N$ the number of bases related to the ambient noises and $K_S$ the number of bases related to the biomedical sounds. In this stage, the described decomposition model (see Equation (3)) does not obtain a parts-based objects reconstruction with physical meaning as occurs in real-world. Therefore, this stage cannot distinguish between spectral patterns belonging to biomedical sounds and ambient noises.

- Stage 2. This stage is applied to the external signal $y(t)$. We assume that the external signal is only composed of ambient noises, therefore the goal of this model is to reconstruct the external magnitude spectrogram $\mathbf{Y}$ by using the basis matrix $\mathbf{U}_N$ composed of the spectral patterns that characterize the ambient noises,

$$\mathbf{Y} \approx \hat{\mathbf{Y}} = \mathbf{U}_N\mathbf{H}_N \quad (3)$$

where $\hat{\mathbf{Y}} \in \mathbb{R}_+^{F\times T}$ is the estimated or reconstructed magnitude spectrogram of the external signal. The variable $\mathbf{H}_N \in \mathbb{R}_+^{K_N\times T}$ is the estimated activations matrix of the ambient noises for the external signal. Note that $\mathbf{U}_N$ can be treated as the same matrix previously used in Equation (2).

Specifically, 2C-NMPCF extends the conventional NMPCF to a multichannel scenario sharing the frequency basis matrix $\mathbf{U}_N$ to factorize simultaneously two distinct single-channel magnitude spectrograms $\mathbf{X}, \mathbf{Y}$ as shown in Figure 3. The proposed method 2C-NMPCF allows to factorize jointly $\mathbf{X}$ and $\mathbf{Y}$ so the spectral patterns of the ambient noises, active in both spectrograms, are shared in the same dictionary $\mathbf{U}_N$ since we assume that ambient noises can be considered repetitive sounds that can be simultaneously active in both magnitude spectrograms $\mathbf{X}$ and $\mathbf{Y}$. Contrarily, the dictionary $\mathbf{U}_S$ represents the spectral patterns of the biomedical sounds that only can be found in the internal magnitude spectrogram $\mathbf{X}$.

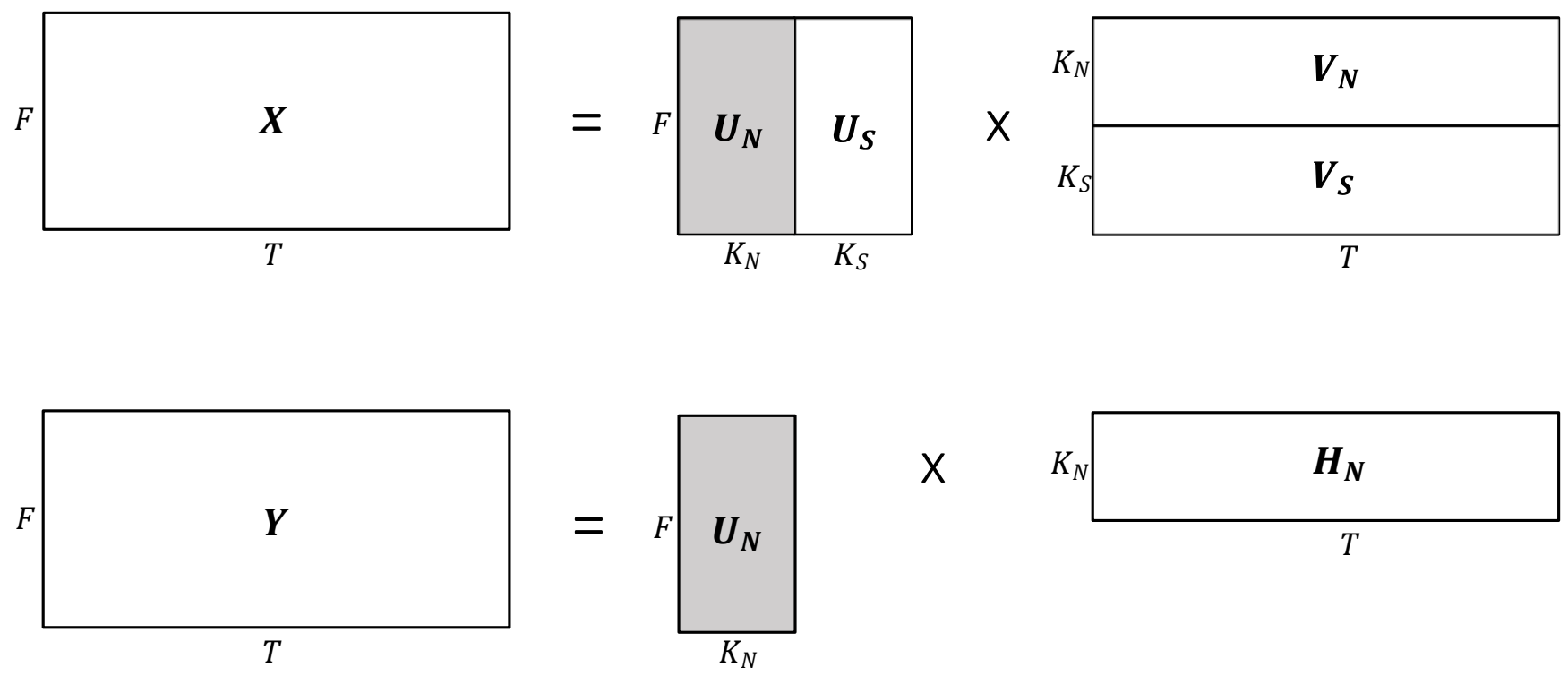

Figure 3: Matrix decomposition based on multichannel NMPCF (2C-NMPCF).

### 2.3.3. Objective Function and Update rules

The objective function of the proposed method 2C-NMPCF that must be performed to minimize the residuals of the two previous models, see Equations (2)-(3), is detailed as follows,

$$\Gamma = D_{KL}\left(\mathbf{X}|\hat{\mathbf{X}}\right) + \lambda D_{KL}\left(\mathbf{Y}|\hat{\mathbf{Y}}\right) \tag{4}$$

where the parameter $\lambda$ controls the relative importance between the internal and the external magnitude spectrogram. So, the contribution of the magnitude spectrogram $\mathbf{Y}$ is greater when the parameter $\lambda$ increases. In this paper, the Kullback–Leibler divergence, see Equation (5), has been used to calculate the signal reconstruction error for the internal spectrogram $D_{KL}\left(\mathbf{X}|\hat{\mathbf{X}}\right)$ and the external spectrogram $D_{KL}\left(\mathbf{Y}|\hat{\mathbf{Y}}\right)$. The reason is because this cost function $D_{KL}$ is non-increasing, ensuring the non-negativity of the estimated basis and activations matrices and moreover, several works have demonstrated promising results in the field of biomedical signal processing [7, 11, 52].

$$D_{KL}\left(\mathbf{Z}|\hat{\mathbf{Z}}\right) = \mathbf{Z}\log\frac{\mathbf{Z}}{\hat{\mathbf{Z}}} - \mathbf{Z} + \hat{\mathbf{Z}}, \quad \mathbf{Z} = \{\mathbf{X}, \mathbf{Y}\} \tag{5}$$

From Equation (4), the estimated basis matrices $\mathbf{U}_N$,$\mathbf{U}_S$ and activation matrices $\mathbf{V}_N$,$\mathbf{V}_S$,$\mathbf{H}_N$ can be obtained by applying a gradient descent algorithm based on multiplicative update rules. The multiplicative update rules to learn those matrices can be obtained by taking negative and positive terms of the partial derivative of the cost function $\Gamma$ with respect to $\mathbf{U}_N$,$\mathbf{U}_S$, $\mathbf{V}_N$,$\mathbf{V}_S$ and $\mathbf{H}_N$, respectively,

$$\mathbf{U}_N \leftarrow \mathbf{U}_N \odot \frac{\left(\mathbf{X} \oslash \hat{\mathbf{X}}\right)(\mathbf{V}_N)^T + \lambda\left(\mathbf{Y} \oslash \hat{\mathbf{Y}}\right)(\mathbf{H}_N)^T}{(\mathbf{V}_N)^T + \lambda(\mathbf{H}_N)^T} \tag{6}$$

$$\mathbf{U}_S \leftarrow \mathbf{U}_S \odot \frac{\left(\mathbf{X} \oslash \hat{\mathbf{X}}\right)(\mathbf{V}_S)^T}{(\mathbf{V}_S)^T} \tag{7}$$

$$\mathbf{V}_N \leftarrow \mathbf{V}_N \odot \frac{(\mathbf{U}_N)^T\left(\mathbf{X} \oslash \hat{\mathbf{X}}\right)}{(\mathbf{U}_N)^T} \tag{8}$$

$$\mathbf{V}_S \leftarrow \mathbf{V}_S \odot \frac{(\mathbf{U}_S)^T\left(\mathbf{X} \oslash \hat{\mathbf{X}}\right)}{(\mathbf{U}_S)^T} \tag{9}$$

$$\mathbf{H}_N \leftarrow \mathbf{H}_N \odot \frac{(\mathbf{U}_N)^T\left(\mathbf{Y} \oslash \hat{\mathbf{Y}}\right)}{(\mathbf{U}_N)^T} \tag{10}$$

where 0 is the element-wise multiplication, 0 is the element-wise division and $()^T$ is the transpose operator. The set of activation and basis matrices for both the internal and external magnitude spectrograms is obtained updating the rules detailed in Equations (6)-(10) using an iterative process until the algorithm converges or reaches a maximum number of iterations $M$.

Focusing on the separation process applied to the biomedical sounds and ambient noises present in the internal spectrogram, the estimated magnitude spectrograms $\hat{\mathbf{X}}_N$ and $\hat{\mathbf{X}}_S$ can be obtained from the estimated basis $\mathbf{U}_N$,$\mathbf{U}_S$ and activation matrices $\mathbf{V}_N$,$\mathbf{V}_S$ as follows:

$$\hat{\mathbf{X}}_N = \mathbf{U}_N\mathbf{V}_N \tag{11}$$

$$\hat{\mathbf{X}}_S = \mathbf{U}_S\mathbf{V}_S \tag{12}$$

In order to denormalize the estimated magnitude spectrograms of the internal spectrogram, the matrices $\hat{\mathbf{X}}_N$, $\hat{\mathbf{X}}_S$ are multiplied by the denominator of Equation (1) when $\mathbf{Z} = \mathbf{X}$. To guarantee a conservative strategy in the reconstruction process, the estimated biomedical signal $\hat{s}(t)$ (Equation (14)) is computed by the inverse overlap-add STFT of the element-wise multiplication between the complex spectrogram $\mathbf{X}_c$ and a Wiener mask [11, 7] that represents the relative energy contribution of the biomedical sounds to the energy of the internal signal $x(t)$. The estimated ambient noise signal $\hat{n}(t)$ (Equation (13)) is calculated in a similar way as explained above in Equation (14), but now taking into account that the Wiener mask explains the relative energy contribution of the ambient noise sounds to the energy of the internal signal $x(t)$.

$$\hat{n}(t) = IDFT\left(\mathbf{X}_c \odot \frac{\left|\hat{\mathbf{X}}_N\right|^2}{\left(\left|\hat{\mathbf{X}}_N\right|^2 + \left|\hat{\mathbf{X}}_S\right|^2\right)}\right) \tag{13}$$

$$\hat{s}(t) = IDFT\left(\mathbf{X}_c \odot \frac{\left|\hat{\mathbf{X}}_S\right|^2}{\left(\left|\hat{\mathbf{X}}_S\right|^2 + \left|\hat{\mathbf{X}}_N\right|^2\right)}\right) \tag{14}$$

The pseudo code of the proposed method 2C-NMPCF for the ambient denoising in auscultation is summarized in the Algorithm 1. Although the proposed method can return the estimated biomedical signal $\hat{s}(t)$ and the estimated ambient noise signal $\hat{n}(t)$, only the signal $\hat{s}(t)$ is required for evaluation purposes in this work.

**Algorithm 1** Ambient denoising using 2C-NMPCF

**Require:** $y(t)$, $x(t)$, $K_N$, $K_S$, $\lambda$ and $M$.

1: Compute the normalized magnitude spectrogram $\mathbf{X}$ of the internal signal $x(t)$ using Equation (1).

2: Compute the normalized magnitude spectrogram $\mathbf{Y}$ of the external signal $y(t)$ using Equation (1).

3: Initialize each activation and basis matrix $\mathbf{U}_N$,$\mathbf{U}_S$,$\mathbf{V}_N$,$\mathbf{V}_S$,$\mathbf{H}_N$ with random non-negative values.

4: Update each activation and basis matrix $\mathbf{U}_N$,$\mathbf{U}_S$,$\mathbf{V}_N$,$\mathbf{V}_S$,$\mathbf{H}_N$ using Equations (6)-(10) for the predefined number of iterations $M$.

5: Compute the estimated magnitude spectrograms $\hat{\mathbf{X}}_N$ using Equation (11).

6: Compute the estimated magnitude spectrograms $\hat{\mathbf{X}}_S$ using Equation (12).

7: Denormalize the estimated magnitude spectrograms $\hat{\mathbf{X}}_N$, $\hat{\mathbf{X}}_S$ multiplying by a factor equal to the denominator of Equation (1) when $\mathbf{Z} = \mathbf{X}$.

8: Synthesize the estimated ambient noise $\hat{n}(t)$ using the Equation (13).

9: Synthesize the estimated biomedical signal $\hat{s}(t)$ using the Equation (14).

**return** $\hat{s}(t)$

*2.3.4. Improving the sound quality of biomedical signals by means of an incremental algorithm based on 2C-NMPCF*

The main limitation of the proposal 2C-NMPCF is related to the objective function (see Equation (4)) used to minimize the residuals of the ambient noise. Specifically, the iterative process based on the multiplicative update rules that obtains the denoised biomedical basis and activation matrices is applied until the convergence of 2C-NMPCF into a local minimum after $M$ iterations. For this reason, 2C-NMPCF by itself is not able to extract all spectral patterns associated to ambient noises. To overcome the previous limitation, we propose an incremental algorithm that runs 2C-NMPCF more than once improving the estimated biomedical signal $\hat{s}_i(t)$ obtained in the incremental iteration $i$ by removing additional spectral content associated to ambient noise that 2C-NMPCF was not able to remove in the previous incremental iteration $i-1$. In general considering the iteration $i$, the internal signal $x_{i+1}(t)$ of the next incremental iteration $i+1$ is the estimated biomedical signal $\hat{s}_i(t)$ obtained in the current incremental iteration $i$, that is, $x_{i+1}(t) = \hat{s}_i(t)$. Note that in the first incremental iteration $i=1$, $x_1(t)=x(t)$. However, the external signal $y(t)$ is fixed for all incremental iterations since we assume that the signal $y(t)$ is only composed by ambient noises (no biomedical sounds are active). The flowchart of the proposed incremental algorithm based on 2C-NMPCF is shown in Figure 4.

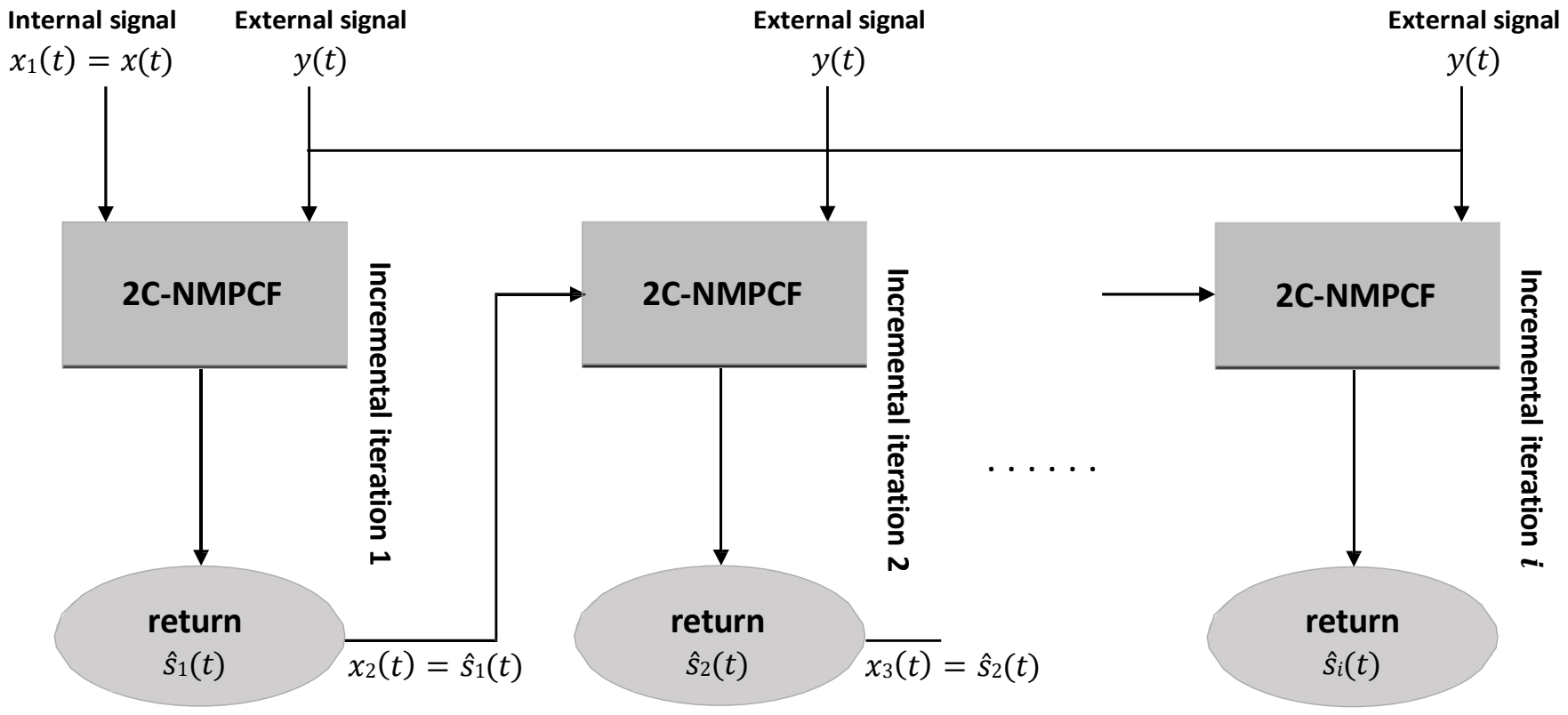

Figure 4: Flowchart of the proposed incremental algorithm based on 2C-NMPCF.

We assume the following assumptions in order to justify our incremental proposal based on 2C-NMPCF: (i) the objective function $\Gamma$ would converge into a better local minimum at each incremental iteration since it would find remainder spectral patterns of ambient noise that have not been extracted in the previous iteration $i-1$ but they are still being repeated in both the internal $\mathbf{X}_i$ and the external $\mathbf{Y}$ magnitude spectrograms in the current incremental iteration $i$; (ii) 2C-NMPCF will remove most of the ambient noise while preserving the content of the biomedical sounds until an optimal number of incremental iterations $i = i_o$. From this optimal iteration $i = i_o$, our proposal can continue to eliminate hidden patterns of ambient noise that are still active at the expense of eliminating also spectral content related to biomedical signals. Summarizing, this incremental approach attempts to maintain most of the biomedical content $\hat{s}_{i_o}(t)$ removing most of the ambient noise through the incremental iterations. An illustrative example of the performance of the proposed incremental approach is shown in Figure 5.

## 3. Evaluation

### *3.1. Metrics*

To evaluate the quality of the biomedical signals estimated by the proposed method, the BSS EVAL toolbox [53, 54] has been used because it proposes a set of metrics, widely applied in the field of sound source separation [11, 7] and background noise removal [55], that quantify the quality of the sound separation between the original biomedical signal and its estimation. Two metrics, measured in dB, are used as occurs in [56, 57]: (i) Source to Distortion ratio (SDR) measures the overall quality of the estimated biomedical signal; and (ii) Source to Interference ratio (SIR) measures the presence of ambient noise in the estimated biomedical signal. Higher values of these ratios indicate better sound quality of the estimated biomedical signal.

In this paper, the optimization and testing results have been obtained calculating both SDR and SIR median values [58]. These results do not show the absolute SDR and SIR values obtained from the estimated biomedical signal $\hat{s}_i(t)$ but the SDR and SIR improvement comparing $\hat{s}_i(t)$ and the original internal signal $x(t)$.

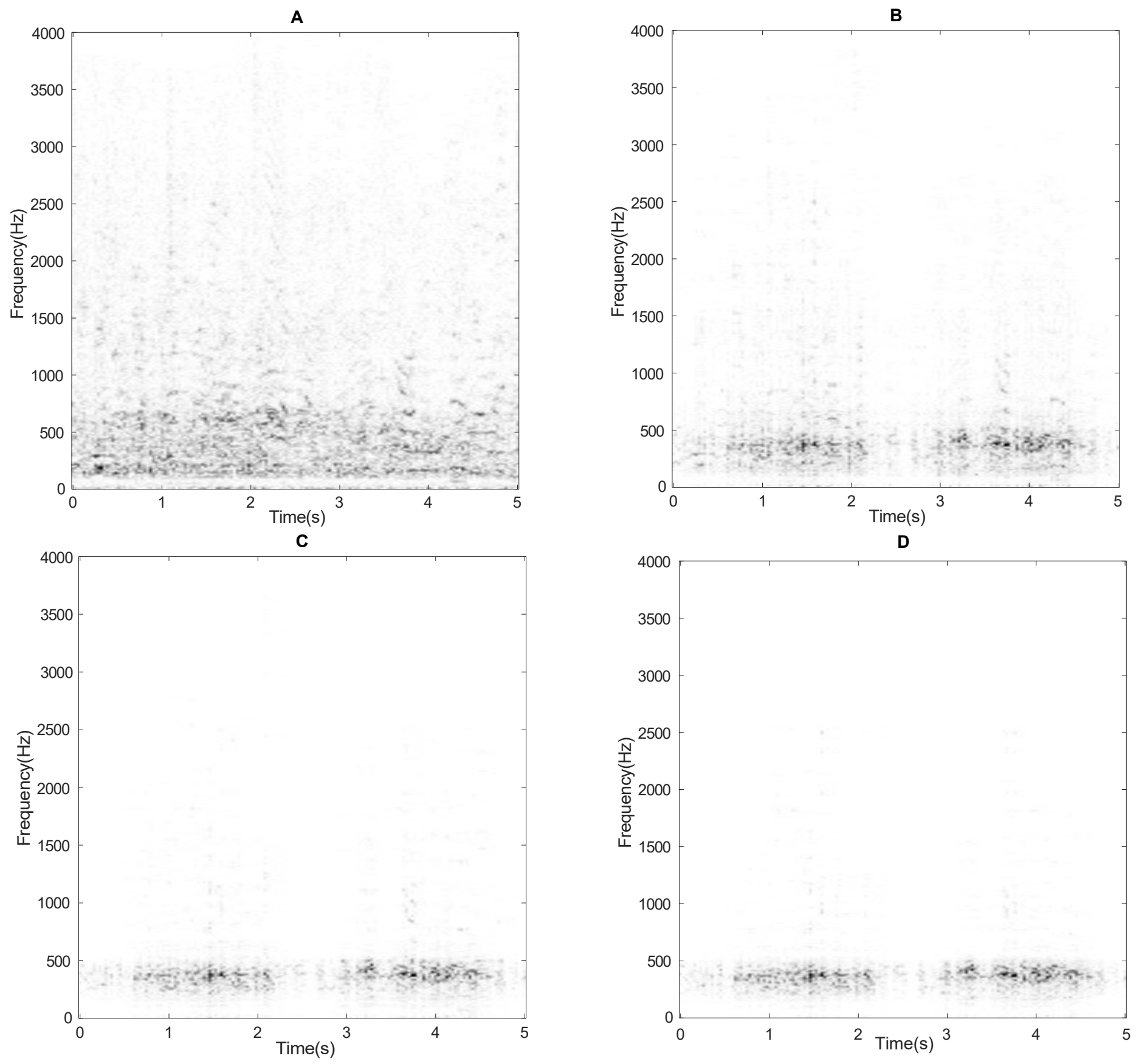

Figure 5: **A**) Magnitude spectrogram $\mathbf{X}$ of the internal signal previously shown in Fig. 1 (Right); Some of the estimated biomedical magnitude spectrograms $\hat{\mathbf{X}}_S$ provided by the incremental algorithm based on 2C-NMPCF through the incremental iterations $i$: **B**) $i = 1$, **C**) $i = 2$ and **D**) $i = 3$. Here, it can be observed that the estimated biomedical spectrogram is refined after each incremental iteration by removing spurious ambient noise that are still active in the previous incremental iteration maintaining most of the biomedical spectral content.

As occurs in [24], two metrics related to speech intelligibility (SI) are added to the objective assessment of the proposed method, which have been previously used in the field of ambient noise suppression in lung auscultation [24, 25]: (i) Normalized-Covariance Measure (NCM); and (ii) Coherence Speech Intelligibility Index (CSII). Specifically, a three-level CSII approach is used by dividing the signal into three amplitude regions: low ($CSII_{low}$), mid ($CSII_{mid}$) and high ($CSII_{high}$). In this work, each metric was computed between the original biomedical signal $s(t)$ and the estimated biomedical signal $\hat{s}(t)$. Note that higher values of these metrics indicate better sound quality of the estimated biomedical signal. Finally, more details of these metrics can be found by the reader in [24, 25, 59].

### *3.2. Setup*

Because most of the spectral content both the biomedical signals [45, 46, 48, 49] and the ambient noise [23, 60] is concentrated below 4 KHz, in this work, a sampling rate equals $f_s = 8$ KHz has been used as occurs in [24].

A preliminary study showed that the following parameters for time-frequency representation provide the best trade-off between the separation performance and the computational cost: a Hamming window with $N = 512$ samples length (64ms) and 50% overlap; and a discrete Fourier transform using 2N points similarly as in [11, 52]. Furthermore, the convergence of the proposed method was empirically observed after 50 iterations, so the parameter $M$ is fixed to $M = 50$.

Finally, note that the performance of the proposed method depends on the initial values with which the basis matrices $\mathbf{U}_S$, $\mathbf{U}_N$ and the activation matrices $\mathbf{V}_S$, $\mathbf{V}_N$, $\mathbf{H}_N$ have been initialised. Although the obtained results are not dispersed and keep the same behavior, in order to overcome this issue, we have run the proposed method three times for each mixture and the results shown in this paper have been calculated using the median values as previously mentioned.

### *3.3. Results*

In this section, experimental results related to the optimization and testing are detailed.

#### *3.3.1. Optimization results*

Several parameters must be fitted to obtain the best performance of the proposed method in the removal of ambient noise. Four parameters have been evaluated using the database $D_O$: (i) The number of biomedical bases $K_S$ used to characterise the spectral content of the biomedical signal $s(t)$, specifically, $K_S \in [16, 32, 64, 128, 256]$; (ii) The number of ambient noise bases $K_N$ used to characterise the spectral content of the ambient noise $n(t)$, specifically, $K_N \in [16, 32, 64, 128, 256]$; (iii) The value $\lambda$ to balance the importance of the internal $\mathbf{X}$ and external $\mathbf{Y}$ magnitude spectrograms in the co-factorization process. In this case, $\lambda \in [0.01, 0.1, 1, 10, 25, 50, 100, 250, 500, 1000]$; (iv) The number of incremental iterations $i$ applied to 2C-NMPCF.

The optimization process is composed of two steps:

1. Step I. Optimize the three parameters $K_S, K_N, \lambda$ in order to reach the greater median of the SDR improvement when applying 2C-NMPCF considering all the types of ambient noises and SNR previously mentioned.
2. Step II. Once the parameters of 2C-NMPCF have been optimized, it must be obtained the optimal number of incremental iterations $i = i_o$ to achieve the best performance of the proposed method (see Figure 4).

Figure 6 shows the median of the SDR improvement analyzing the search space derived from the parameters $\lambda$, $K_S$ and $K_N$. Results indicate that the proposed method provides the best denoising performance, by means of the maximum median value of the SDR improvement, using the optimal parameters $K_N$=256 and $K_S$=16. These optimal values demonstrate that ambient noise requires a greater number of spectral patterns compared to biomedical sounds due to their greater spectral diversity. The analysis of different lung and heart signals indicates that the spectral modeling of these biomedical sounds is simpler and therefore needs a smaller dictionary of bases since lung sounds could be factorize by a low-rank decomposition using broadband spectral patterns that show temporal and spectral smoothness. However, heart sounds could be modeled as low-frequency pulses located in regular intervals in time.

Figure 7 shows the median of the SDR and SIR improvement using the previous optimal values of the parameters $K_S$ and $K_N$ (that is, $K_S$= 16 and $K_N$=256). It can be confirmed that giving importance to the sharing of spectral bases in the joint factorization process finds a better local minimum in the factorization process, since ambient noise clearly reveals its simultaneous presence both in the internal and external signal spectrogram. Results report a significant and stable SDR and SIR improvement equals 14 dB and 19.5 dB using $\lambda \geq 10$. For this reason, we have chosen the optimal parameter $\lambda$=10.

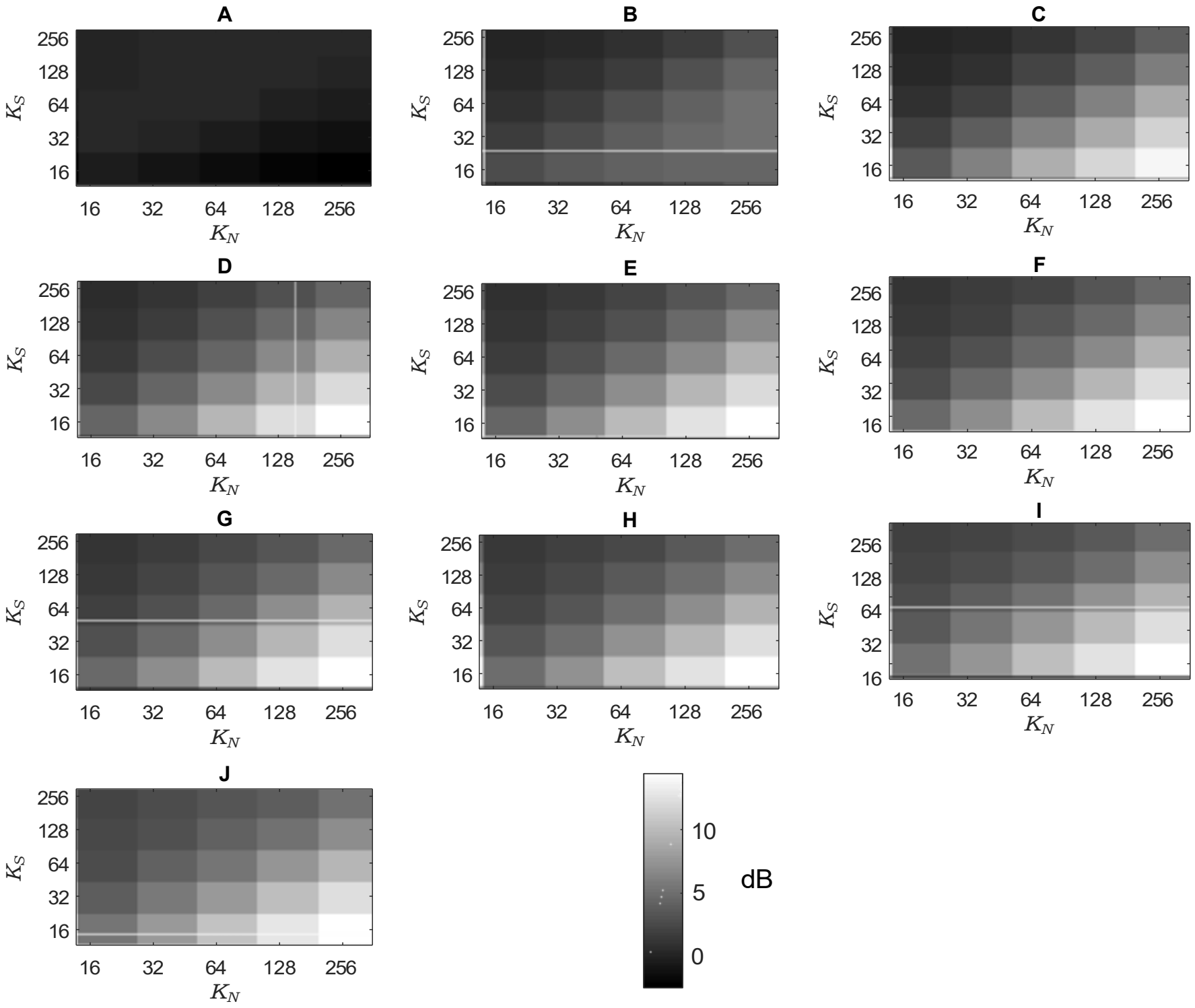

Figure 6: Median values of the SDR improvement (dB) evaluating $D_O$ for the following $\lambda$ values: $\lambda$ =0.01 (A), $\lambda$ =0.1 (B), $\lambda$ =1 (C), $\lambda$ =10 (D), $\lambda$ =25 (E), $\lambda$ =50 (F), $\lambda$ =100 (G), $\lambda$ =250 (H), $\lambda$ =500 (I) and $\lambda$ =1000 (J).

Figure 8 depicts the optimal number of incremental iterations $i$ of the proposed method using the previous optimal parameters $K_S$, $K_N$ and $\lambda$. It is observed that both SDR and SIR improvement increases sharply when applying 2C-NMPCF in the second incremental iteration ($i$ =2), higher increase for SIR compared to SDR. In the third incremental iteration ($i$ =3), the SDR im-provement increases slightly and then starts to decrease gradually while the SIR improvement continues to grow. Experimental results indicate that ambient noise continues to be suppressed at the expense of starting to remove biomedical spectral content when $i > 3$. For this reason, the optimal number of incremental iterations has been set at $i_o = 3$ with the aim of providing the greatest suppression of ambient noise while maintaining most of the biomedical content, being $i_o$ the iteration in which the maximum SDR improvement is obtained.

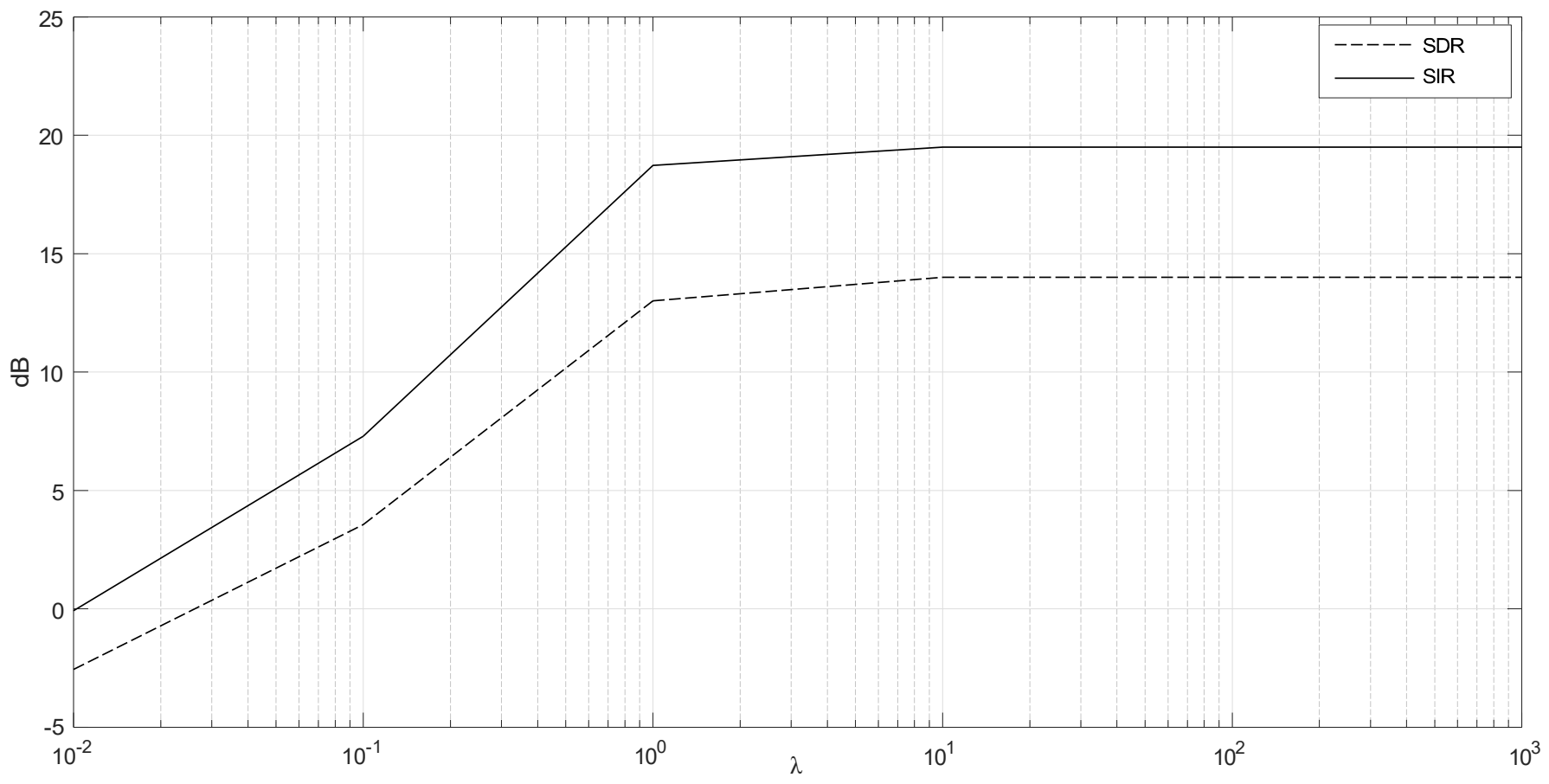

Figure 7: Median values of the SDR (dashed line) and SIR (solid line) improvement of the proposal algorithm evaluating $D_O$, keeping fixed $K_S$=16 and $K_N$=256 varying the parameter $\lambda$.

### *3.3.2. Objective results simulating an ideal scenario*

This section evaluates the ambient denoising performance of the proposed method simulating an ideal scenario since neither the effects of propagation on the patient's body material nor the acoustics of the room are considered active (these effects will be analyzed in section 3.3.3.).

Figure 9 shows SDR and SIR improvement comparing the behaviour of the ambient noise removal evaluating the database $D_T$ for the proposed method (2C-NMPCF) and the baseline methods MSS and NLMS. Hereinafter, the SDR and SIR improvement associated to each method are represented by $SDR_P$ and $SIR_P$ (the proposed method), $SDR_M$ and $SIR_M$ (MSS) and finally, $SDR_N$ and $SIR_N$ (NLMS). Each box represents 250 data points, one for each recording of the database $D_T$. The lower and upper lines of each box show the 25th and 75th percentiles. The line in the middle of each box represents the median value. The diamond in the center of each box represents the average value. The lines extending above and below each box show the extent of the rest of the samples, excluding outliers. Outliers are defined as points that are over 1.5 times the interquartile range from the sample median, which are shown as crosses.

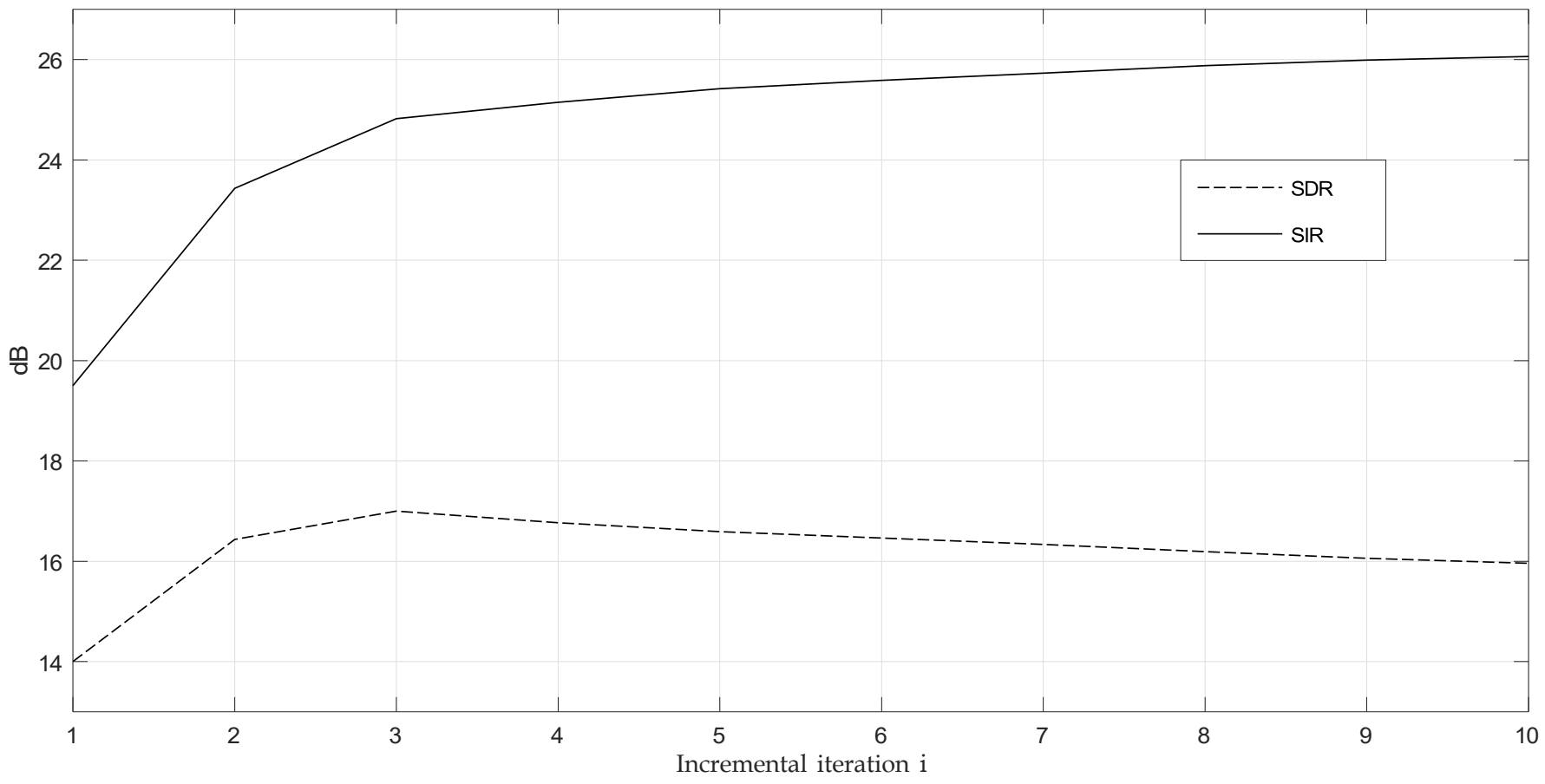

Figure 8: Median values of the SDR (dashed line) and SIR (solid line) improvement with the number of incremental iterations $i$ evaluating $D_O$, keeping fixed ($K_S$, $K_N$, $\lambda$)=(16, 256, 10).

Figure 9A shows that the proposed method outperforms MSS and NLMS, in terms of SDR, in most of the SNR, obtaining the best average performance when all SNRs are taken into account. It can be seen how the proposed method reports the most robust SDR performance compared to MSS and NLMS, mainly in high noise environments (SNR∈[-20 dB,-10 dB]), by means of a stable SDR trend in this SNR range. It suggests that the multichannel co-factorization approach is less dependent on the ratio of ambient noise to biomedical content than the other MSS and NLMS methods. Although MSS shows an increasing trend in SDR improvement stabilizing its denoising results for SNR≥-10dB, these results are not satisfactory enough to consider MSS as a competitive method with respect to the proposed method or NLMS since the same does not happen in terms of SIR improvement (see Figure 9B) where MSS provides $SIR_M \leq 14$dB in high noise environments compared mainly to the proposed method.

Figure 9B indicates that although the proposed method and NLMS obtain a very similar performance considering the average of all SNRs, NLMS is slightly better in SNR $\leq$ -10 dB at the expense of losing a large amount of biomedical content, a fact that does not occur with the proposed method. This behavior shown by NLMS is confirmed by its SDR and SIR reduction as SNR increases. Focusing on the proposed method, the stable SDR and SIR trends regardless of the SNR evaluated demonstrate that our incremental approach is a more reliable feature to remove ambient noise because: i) MSS provides a satisfactory performance assuming a distortion of biomedical sounds at low frequencies and penalizing the occurrences of noise with strong energy in high spectral bands [24]; and ii) the proposed method assumes ambient noise as a repetitive sound event found in both the internal and external spectrograms, so it does not depend on any temporal misalignment between the internal and external signal as occurs with NLMS.

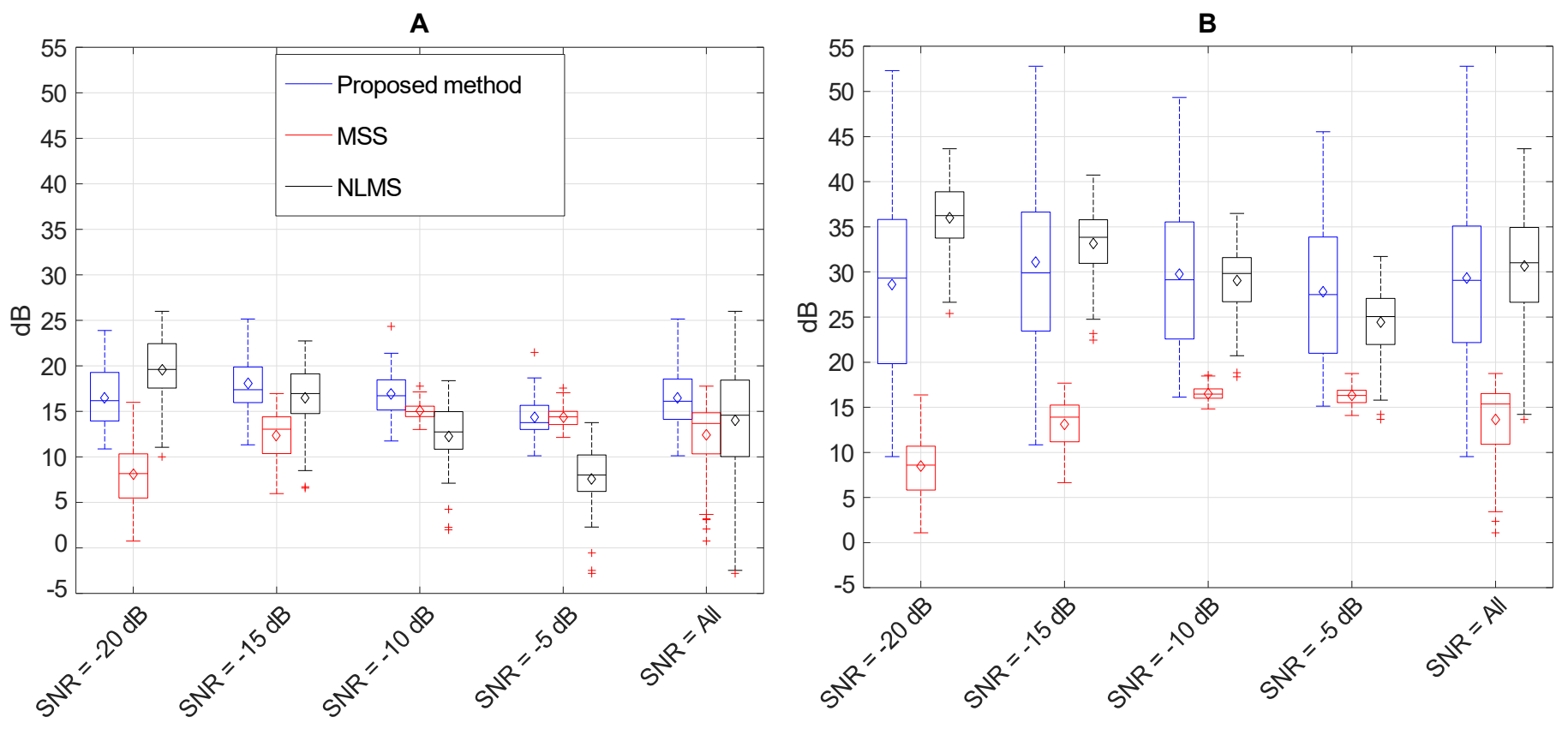

Figure 9: SDR (A) and SIR (B) improvement considering all the noises (Ambulance Siren, Baby Crying, Babble, Car and Street) for each SNR from database $D_T$. Each box represents the denoising performance results for each method evaluated. The color of the legend, associated to each evaluated method shown in the subfigure 9.A, refers to the same methods for all subfigures.

Figure 10 shows a detailed analysis of the denoising performance of the proposed method considering each particular type of ambient noise evaluated previously in Figure 9. Each box represents 50 data points, one for each recording of the database $D_T$. Highlight the stable SDR and SIR trends regardless of the type of noise and level SNR evaluated shown by the proposed method unlike what happens with MSS and NLMS.

- Ambulance siren noise: Figures 10A and 10B show that although NLMS only obtains better SDR results compared with the proposed method in high noisy scenarios (SNR≤-10 dB), the denoising performance of both methods are still competitive averaging all evaluated SNRs. Nevertheless, it is interesting that the SDR improvement of the proposed method outperforms NLMS in acoustic scenarios in which SNR≥-10 dB because it is a frequent acoustic scenario that can be found around an auscultation room located inside a health center.

- Baby crying noise: Figures 10C and 10D show that the proposed method significantly improves the SIR denoising performance compared with MSS and NLMS keeping most of the biomedical content as shown SDR results, an interesting advantage particularly in high noisy environments. Results seem to suggest that the strong harmonic structure contained into this type of ambient noise facilitates the sound separation between noise and biomedical sounds since the more dissimilar the spectral patterns of the noise and the biomedical signal, the better the noise suppression performance of the proposed method.

- Babble noise: Figures 10E and 10F indicate that NLMS obtains the best SDR and SIR improvements evaluating this particular ambient noise but showing a decreasing trend as the SNR increases, unlike the other evaluated methods. Experimental results indicate that the spectrum of the Babble noise and the biomedical signal, mainly lung sounds, is more similar compared to the above ambient noises (ambulance siren and baby crying noise) so, the greater are the spectral differences between the target sounds and the noise, the better is the performance of the proposed method since the repetitive behaviour of the ambient noise in the co-factorization process is more easy to suppress.

- Car noise: Figures 10G and 10H indicate that the proposed method obtains the best ambient denoising performance compared to MSS and NLMS since it achieves the best biomedical audio quality by maximizing the amount of suppressed ambient noise at the expense of preserving most of the biomedical content related to the estimated biomedical signal of interest.

- Street noise: Figures 10I and 10J show that NLMS preserves higher amount of biomedical content in high noisy environments (SNR$\leq$-10 dB) compared to the proposed method and MSS. However, SDR and SIR results indicate that the proposed method and MSS are competitive compared to NLMS. Moreover, NLMS reports the first SIR ranking position in all SNR scenarios. Finally, the proposed method outperforms, in terms of SDR results, MSS and NLMS in the remainder SNRs.

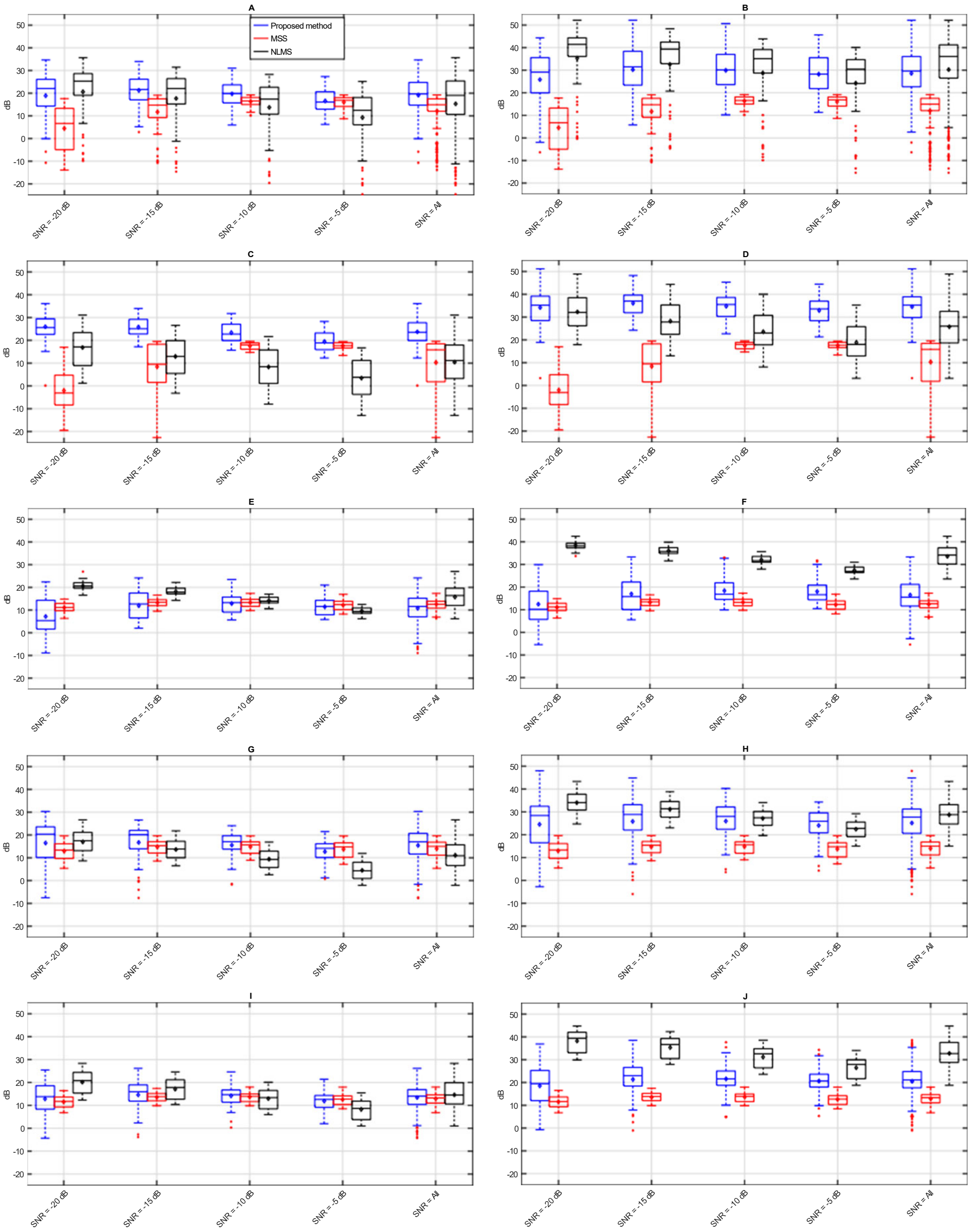

Figure 10: SDR and SIR improvement results provide by the proposed method and the baseline methods (MSS and NLMS) evaluating $D_T$ and each type of ambient noise along SNR: Ambulance Siren (A and B), Baby crying (C and D), Babble (E and F), Car (G and H) and Street (I and J). Each box represents the denoising performance results for each method evaluated. The color of the legend, associated to each evaluated method shown in the subfigure 10.A, refers to the same methods for all subfigures.

Figure 11 shows the effect of inserting a time delay between the internal and external signals. The purpose of the delay is to simulate the time processing that takes the digital stethoscope to apply filtering, artifacts removal and other signal processing operations [61]. It can be observed how the delay affects to the computation of the median value of the SDR and SIR improvement evaluating all the previous ambient noises and SNRs. Results confirm that the most remarkable advantage of the proposed method is its robustness with the variation of the delay. In fact, the proposed method shows a stable behavior in relation to the delay variation between the internal and external signals used in the co-factorization, in contrast to a higher dependence of MSS compared to the proposed method and a higher dependence of NLMS compared to MSS. The ambient denoising performance between the proposed method, MSS and NLMS is accentuated when the delay is active since SDR and SIR results obtained by MSS and NLMS are reduced as the delay increases. Comparing with MSS and NLMS, the proposed method obtains an improvement of 13.1 dB and 16.5 dB in terms of SDR, and 21 dB and 21.5 dB in terms of SIR applying a delay equals to 25 milliseconds. As previously mentioned, results confirm the proposed method as a more appropriate approach to remove the ambient noise because the multichannel co-factorization is based of the noise modelling as repetitive spectral patterns that can be found in any time of the internal and external spectrograms, avoiding errors due to presence of temporal misalignment between both spectrograms.

Figure 12 shows the metrics NCM, $CSII_{low}$, $CSII_{med}$ and $CSII_{high}$ results comparing the performance of the ambient noise removal evaluating the database $D_T$ for the proposed method, MSS and NLMS. The higher metric value is obtained, greater acoustic similarity between the estimated biomedical signal and the original biomedical signal is with respect to their sound contents as occurs in [24]. Specifically, each box represents 250 data points, one for each recording of the database $D_T$. Figure 12 reports that, in general, the best average ambient denoising results are obtained by NLMS. However, it is observed that the average ambient noise suppression achieved by the proposed method can be considered very similar to MSS in most cases of SNR. Broadly, results obtained by each evaluated method improve the acoustic quality of biomedical sounds by eliminating ambient noise that typically hinders clinical examination. Finally, it can be seen how the above results fall as the SNR decreases, similar to what happens in the real world because it is more difficult to hear biomedical sounds in those acoustic scenarios where biomedical sounds are barely audible due to high ambient noise levels.

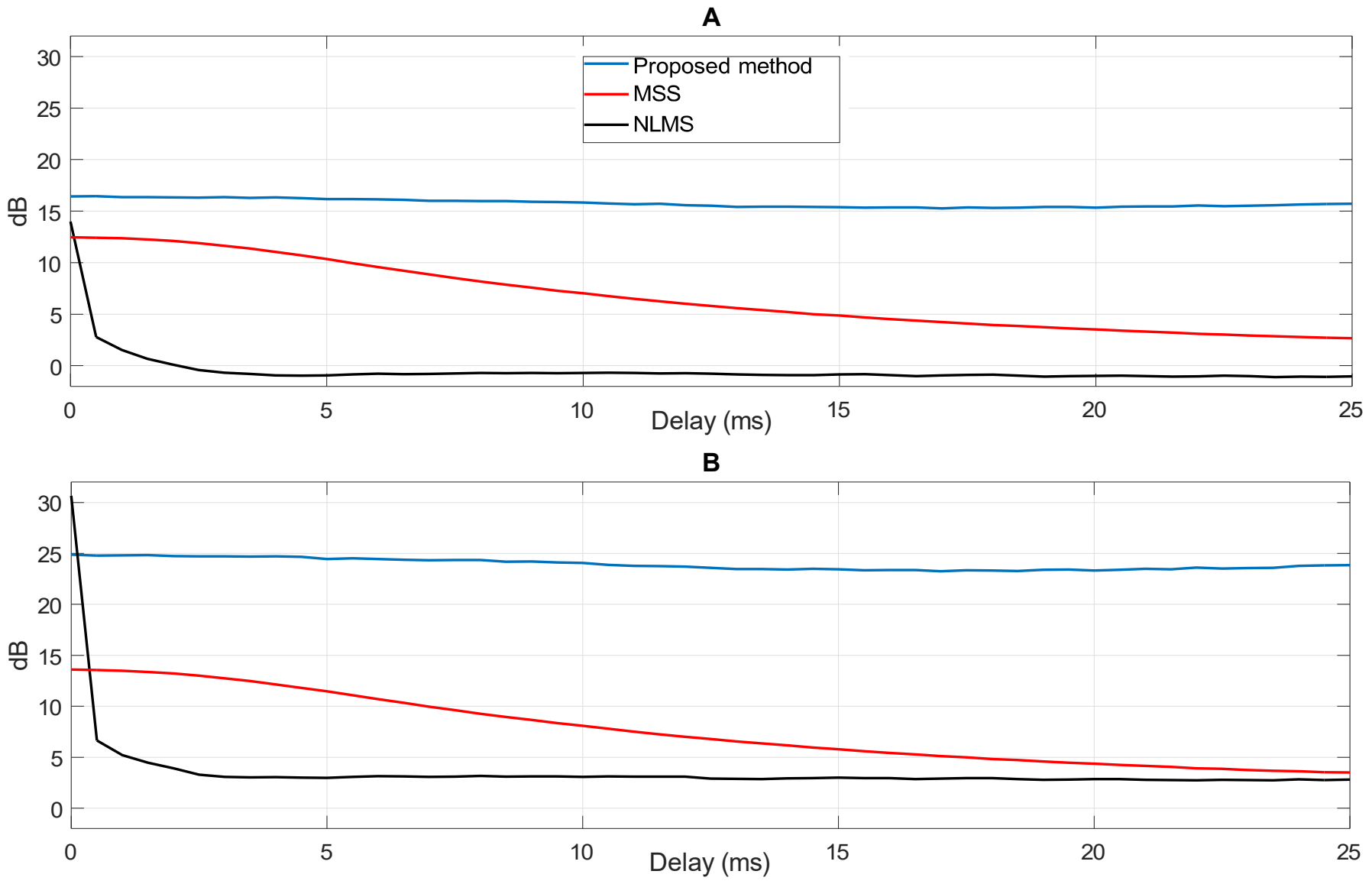

Figure 11: Median values of the SDR (A) and SIR (B) improvement analyzing all the noises (Ambulance Siren, Baby Crying, Babble, Car and Street) and SNRs (-20 dB, -15 dB, -10 dB, -5 dB) varying the delay between the internal and external spectrograms in the database $D_T$. The color of the legend, associated to each evaluated method shown in the subfigure 11.A, refers to the same methods for all subfigures.

### 3.3.3. *Objective results simulating a real scenario*

This section assesses the ambient denoising performance of the proposed method simulating a real scenario where both the propagation in the patient's body material and the acoustics of the room have been considered active.

From the noise recordings belonging to the database $D_N$ (see section 2.1),

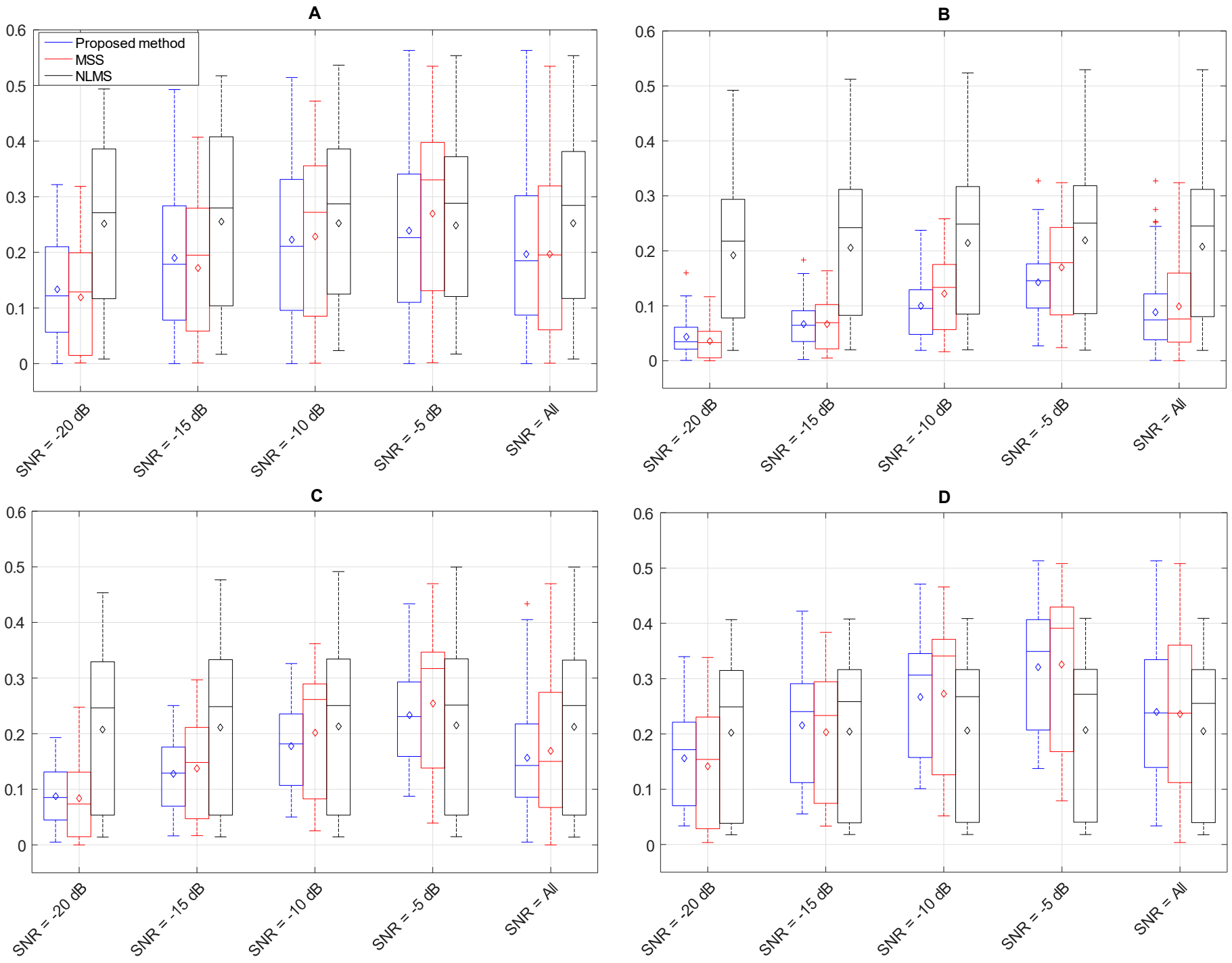

Figure 12: NCM (A), $\mathrm{CSII}_{low}$ (B), $\mathrm{CSII}_{med}$ (C) and $\mathrm{CSII}_{high}$ (D) results considering all the noises (Ambulance Siren, Baby Crying, Babble, Car and Street) for each SNR from database $D_T$. Each box represents the denoising performance results for each method evaluated. The color of the legend, associated to each evaluated method shown in the subfigure 12.A, refers to the same methods for all subfigures.

we have selected those indoor and outdoor ambient noises that can be heard surround a medical consultation room located in a health center: ambulance siren [35, 36], baby crying [37], babble (people speaking) [38, 39] and street (car passing by, car engine running, car idling, bus, truck, children yelling, people talking, workers on the street) [41, 42], that is, a total of 120 single-channel recordings of noises (30 recordings per type of noise) lasting 5 seconds each noise recording as occurred in section 2.1.

The database $D_F$ has been created by modifying the database $D_T$ in order to evaluate the ambient denoising performance of the proposed method taking into account the effects of propagation on the materials of the patient's body and the acoustics of the room. Specifically, two impulse responses, from the open-access dataset [62, 63] and measured with an adult human, have been applied to the previous database $D_T$. The first impulse response $h_H(t)$ has been calculated locating the microphone in the position 33 [62] because it can be considered a correct placement to replicate a heart auscultation in real conditions. The second impulse response $h_L(t)$ has been calculated locating the microphone in the position 55 [62] because it can be considered a correct placement to replicate a lung auscultation in real conditions. Both impulse responses have been resampled at 8kHz and calculated using a sound source located behind the chest of the human. More details related to the propagation on the patient's body material can be found in [62]. In order to simulate the acoustics of the room, we have assumed a standard room in the Hospital Universitario de Jaen (Spain) with dimensions of 7m of large, 4m of width and 2.7m of height using the well-known image method [64] and the MATLAB RIR generator[1]. Specifically, the room impulse response $h_R(t)$ has been designed assuming a moderate reverberation time $RT_{60}$ of 0.4 seconds. The sensor is an omnidirectional microphone placed at the center of the room and the sources are placed outside the room, one in the waiting room and the other outdoors. As occurs with the database $D_T$, several SNR have been applied in the mixing process to create the database $D_F$ in order to evaluate high noisy environments. In this way, the databases $D_{F_{-20}}$ (SNR=-20 dB), $D_{F_{-15}}$ (SNR=-15 dB), $D_{F_{-10}}$ (SNR=-10 dB) and $D_{F_{-5}}$ (SNR=-5 dB) refer to the same database $D_F$ but using different SNR between biomedical and ambient noise recordings by means of the mixing process shown in Figure 13. The mixing process is performed as follows: i) each ambient noise signal $n_i(t)$ (from the database $D_N$) is convoluted with the impulse response $h_R(t)$ to create the external signal $y(t)$. Here, the signal $y(t)$ is only affected by the acoustics of the room; ii) the convolution of the external signal $y(t)$ and the impulse response of the human body, $h_H(t)$ or $h_L(t)$, generates the ambient noise signal $n_{iRB}(t)$. Here, the signal $n_{iRB}(t)$ is affected by both the room acoustics and the patient's body material; iii) the ambient noise signal captured by the stethoscope $n_{iMIX}(t)$ corresponds to the sum of $n_{iRB}(t)$ and $y(t)$, that is, $n_{iMIX}(t)=n_{iRB}(t)+y(t)$; and iv) a desired SNR, in decibels, is obtained between the biomedical signal $s(t)$ and the ambient noise signal $n(t)$. The signal $n(t)$ is calculated weighting $n_{iMIX}(t)$ by the parameter $\alpha = 10^{\frac{(SNRinit-SNR)}{20}}$, being *SNRinit* the initial signal-to-noise ratio between the biomedical signal $s(t)$ and the signal $n_{iMIX}(t)$.

[1]https://www.audiolabs-erlangen.de/fau/professor/habets/software/rir-generator

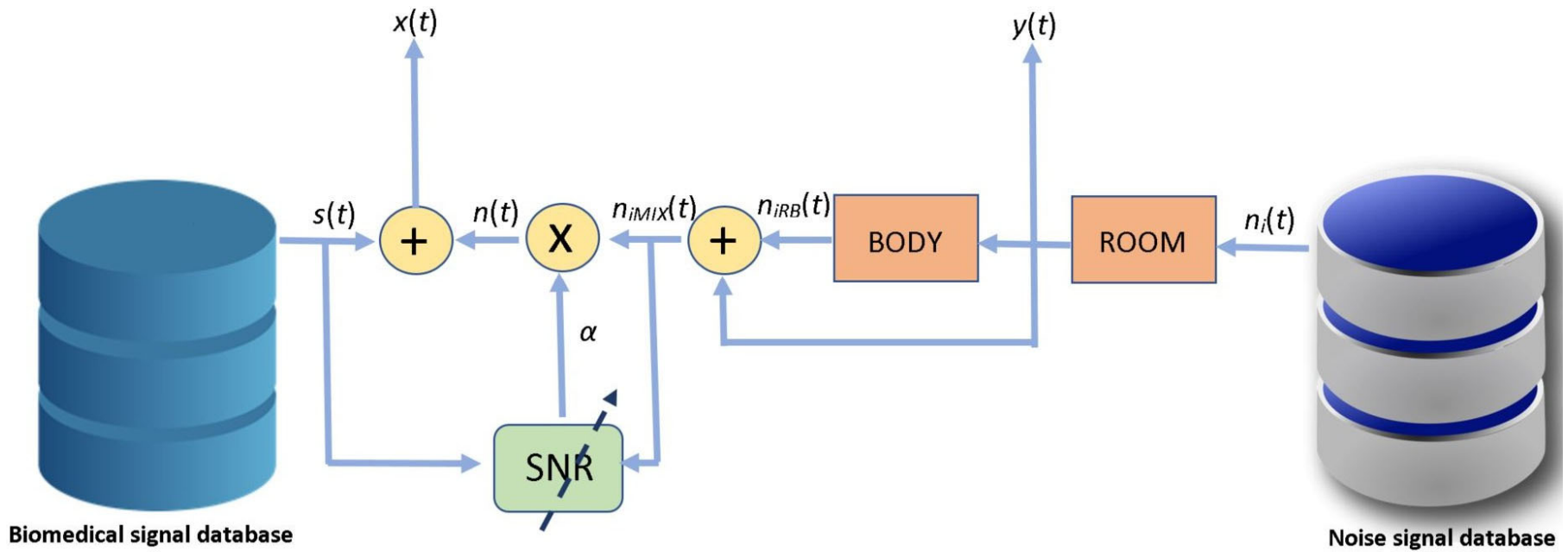

Figure 13: Scheme to simulate a more realistic scenario in which both the effects of propagation on the patient's body material and the acoustics of the room are active.

Figure 14 shows that the proposed method obtains the best average results for SDR (16 dB additional with respect to MSS and 19 dB additional with respect to NLMS) and SIR (23 dB additional with respect to MSS and 13 dB additional with respect to NLMS), significantly outperforming the results provided by MSS and NLMS for each SNR evaluated. Once again, the stable trend of SDR and SIR achieved by the proposed method for each SNR in comparison with the rest of the methods is remarkable. The SDR results indicate that neither MSS nor NLMS can be considered competitive methods in this scenario since the results obtained by both methods yield mean values of SDR $\leq$ 1 dB, which implies a huge loss of biomedical content despite the fact that NLMS gets a great isolated biomedical signal, showing mean values of $SIR_N \in$ [8 dB, 14dB] but lower than those obtained by the proposed method for each SNR. Moreover, the worst performance is provided by MSS in high noisy en-vironments (SNR<-10dB) because the large amount of noise mixed with the biomedical signal hinders optimal MSS performance by avoiding to estimate a correct SNR-dependent factor. Nevertheless, these results again confirm the better performance of NLMS compared to MSS similar to what occurs in Figure 9.

Comparing Figure 9 and Figure 14, it is evidenced the high robustness of the proposed method compared to MSS and NLMS since the average reduction related to the SDR and SIR improvement provided by the proposed method taking into account the effects of this scenario is ≤ 1 dB and ≤ 5 dB while the same SDR and SIR reduction increases to 14 dB and 15 dB for MSS, and 18 dB and 21 dB for NLMS. Figure 14 reports that the estimated denoised spectrogram that models the ambient noise as a repetitive sound event in a cofactorized NMF where simultaneously active repetitive spectral patterns are sought in two different spectrograms provides higher sound quality compared to the spectral subtraction approach or the adaptive filtering technique where the use of temporal information between both spectrograms is essential to correctly recover the target signal.

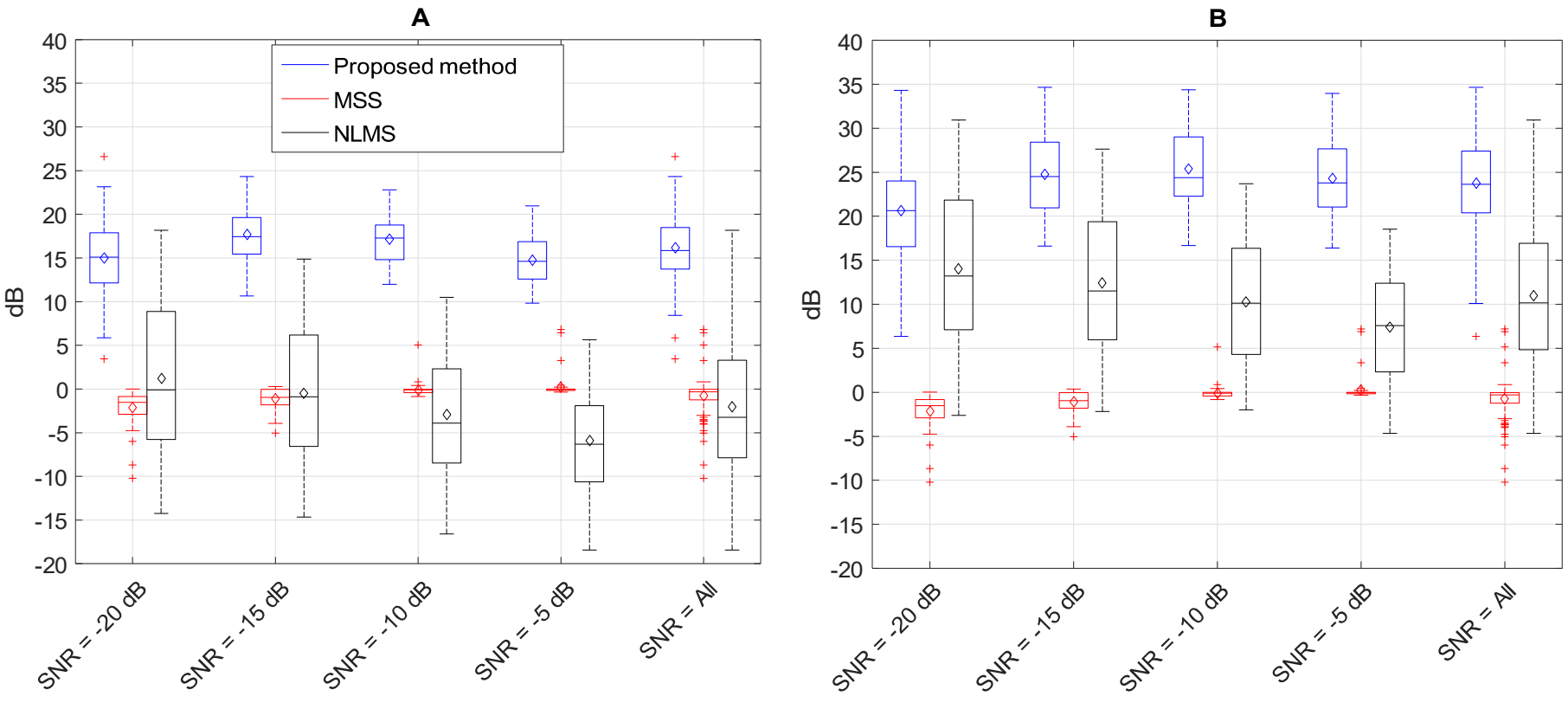

Figure 14: SDR (A) and SIR (B) improvement considering all the noises (Ambulance Siren, Baby Crying, Babble and Street) for each SNR from database $D_F$. Each box represents the denoising performance results for each method evaluated. The color of the legend, associated to each evaluated method shown in the subfigure 14.A, refers to the same methods for all subfigures.

Figure 15 evidences that the proposed method provides the best improvements of SDR and SIR for each type of ambient noise in most SNR by means of maximizing the denoising of ambient noise (highest $SIR_P$) at the expense of minimizing the loss of biomedical energy (highest $SDR_P$). It is also confirmed, as occurs in Figure 10, that the proposed method shows a stable trend in SDR and SIR unlike what happens with the other methods evaluated. Comparing Figure 15 and Figure 10, it can be deduced that the effect of sound propagation through the patient's body material and room acoustics reduces the denoising performance, in terms of SDR and SIR, in all the evaluated methods, however, this reduction is significantly lower in the proposed method compared to MSS and NLMS since the performance of MSS drops sharply in this acoustic environment obtaining $SDR_M$ and $SIR_M$ results $\leq$ 0 dB, being the proposed method and NLMS (only for the case of street noise) the only competitive methods. Again, these results demonstrate what was previously stated in section 3.3.2, that the non-dependence of the temporal information, as occurs in the proposed method, between both spectrograms associated with the internal and external signal achieves a more adequate behavior in the denoising of environmental noise in comparison with those methods (MSS and NLMS) in which the temporal information is used in greater or less way.

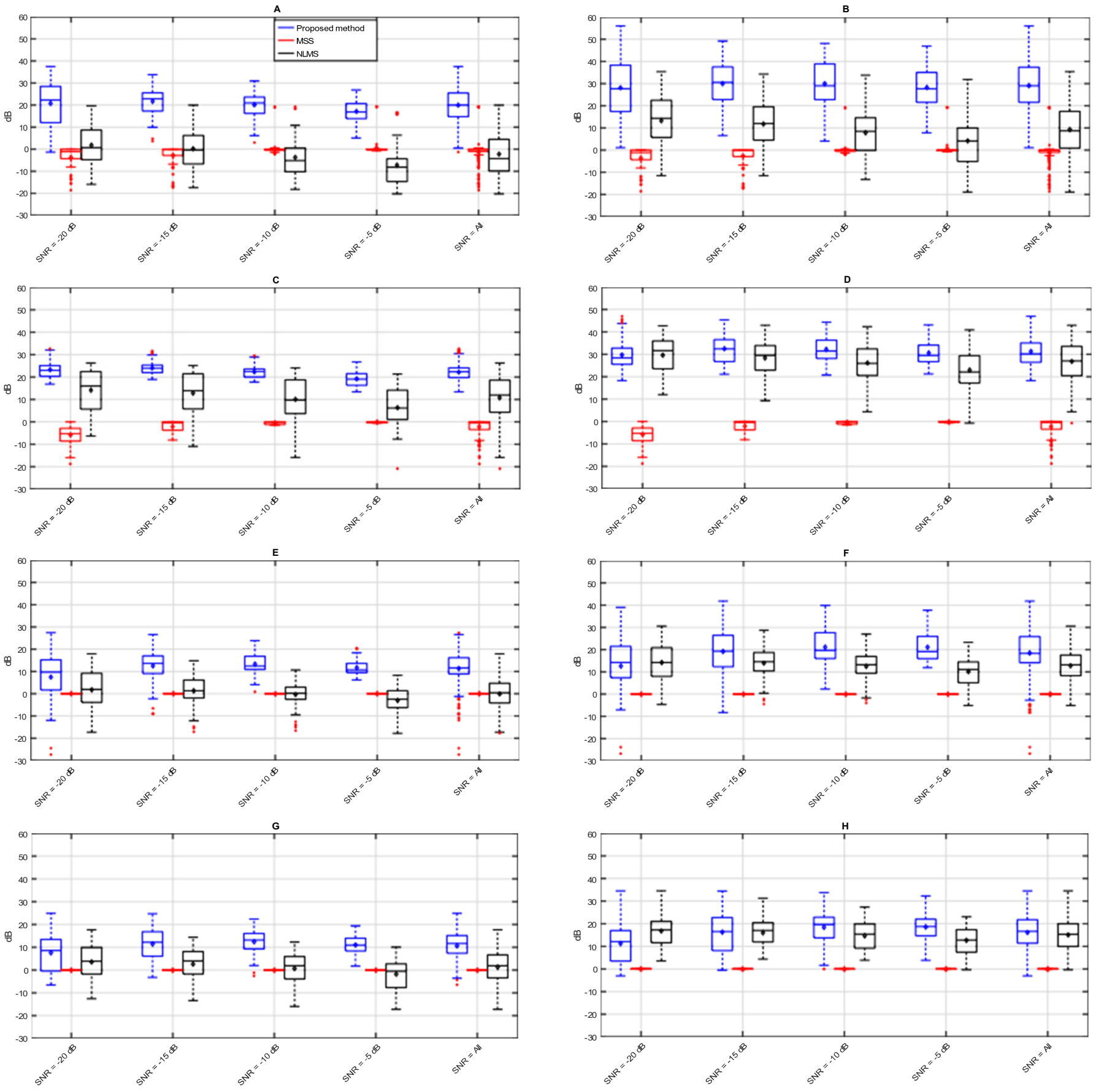

Figure 15: SDR and SIR improvement results provide by the proposed method and the baseline methods (MSS and NLMS) evaluating $D_F$ and each type of ambient noise along SNR: Ambulance Siren (A and B), Baby crying (C and D), Babble (E and F) and Street (G and H). Each box represents the denoising performance results for each method evaluated. The color of the legend, associated to each evaluated method shown in the subfigure 15.A, refers to the same methods for all subfigures.

Figure 16 shows that the proposed method obtains the highest metrics NCM, $CSII_{low}$, $CSII_{med}$ and $CSII_{high}$ for all SNR evaluated increasing the denoising performance as the SNR increases. NLMS is ranked in second position followed by MSS in the last position. The ambient denoising performance provided by MSS (Figure 12) slightly exceeds the proposed method but this performance is not the same when the propagation of the patient's body material and room acoustics are active (Figure 16). Comparing Figure 12 and Figure 16 and unlike what happens with the average values of NLMS, it is observed that the proposed method shows a clear increasing trend in these metrics as the SNR increases. As a result, it seems that the acoustic quality of the biomedical content estimated by the proposed method using these metrics is influenced to a greater extent by the ratio between the biomedical signal and the ambient noise in the input mixture, as opposed to its stable trend previously reported for SDR and SIR. Although the denoising performance provided by MSS and NLMS in Figure 16 drops significantly for all evaluated metrics and SNR compared to the results obtained in Figure 12, the same results obtained by the proposed method are also reduced but this reduction can be considered marginal compared to that suffered by both MSS and NLMS. As a result, it can be deduced that the temporal non-dependence between the internal and external spectrograms in a multichannel co-factorization is a suitable characteristic to remove ambient noise in biomedical sounds mainly considering the effects typically found to model a real scenario (e.g., propagation of the patient's body material, reverberations in the room, ...). These effects drastically reduce the acoustic quality of the extracted biomedical signal as well as the robustness of the method since any temporal misalignment or energy changes caused by impulse responses related to the previous effects imply a relevant signal distortion in the spectrograms, being those approaches based on spectral subtraction (MSS) or adaptive filtering (NLMS) less efficient to minimize the sound interference caused by the previous ambient noises.

*3.3.4. Computational complexity assessment*

In this section, the analysis of the computational complexity of each method evaluated considering the number of elementary operations (multiplications and additions) is detailed in Table 2. The number of operations per second can be obtained taking into account the parameters on which each algorithm used in the experimental results of this work is based: i) the proposed method with $N$=512 samples (Hamming window), $F$=512 bins, $K_N$=256 ambient noise components, $K_S$=16 biomedical components, $M$=50 iterations of the multiplicative update rules, $i_o$=3 incremental iterations and $T$=32 frames per second, implies a total computational cost of $3.32 \times 10^9$ multiplications per second and $1.99 \times 10^9$ additions per second; ii) MSS with $N$=400 samples (Hamming window), $F$=400 bins and $T$=23 frames per second, obtains a total computational cost of $2.78 \times 10^5$ multiplications per second and $4.10 \times 10^5$ additions per second; and iii) NLMS with $L$=10 coefficients of the adaptive filter and $f_S$=8000 samples per second, obtains a total computational cost of $2.48 \times 10^5$ multiplications per second and $2.40 \times 10^5$ additions per second. Results confirm that the proposed method requires a greater number of elementary operations compared to the other evaluated methods, since most of its computational cost is derived from the iterative process based on the multiplicative update rules to estimate the bases and activations matrices used to model the spectro-temporal behavior of biomedical sounds and ambient noise factorized into the NMPCF decomposition in order to extract those spectral patterns active in the magnitude spectrograms corresponding to the internal and external signal (see section 2.3.3).

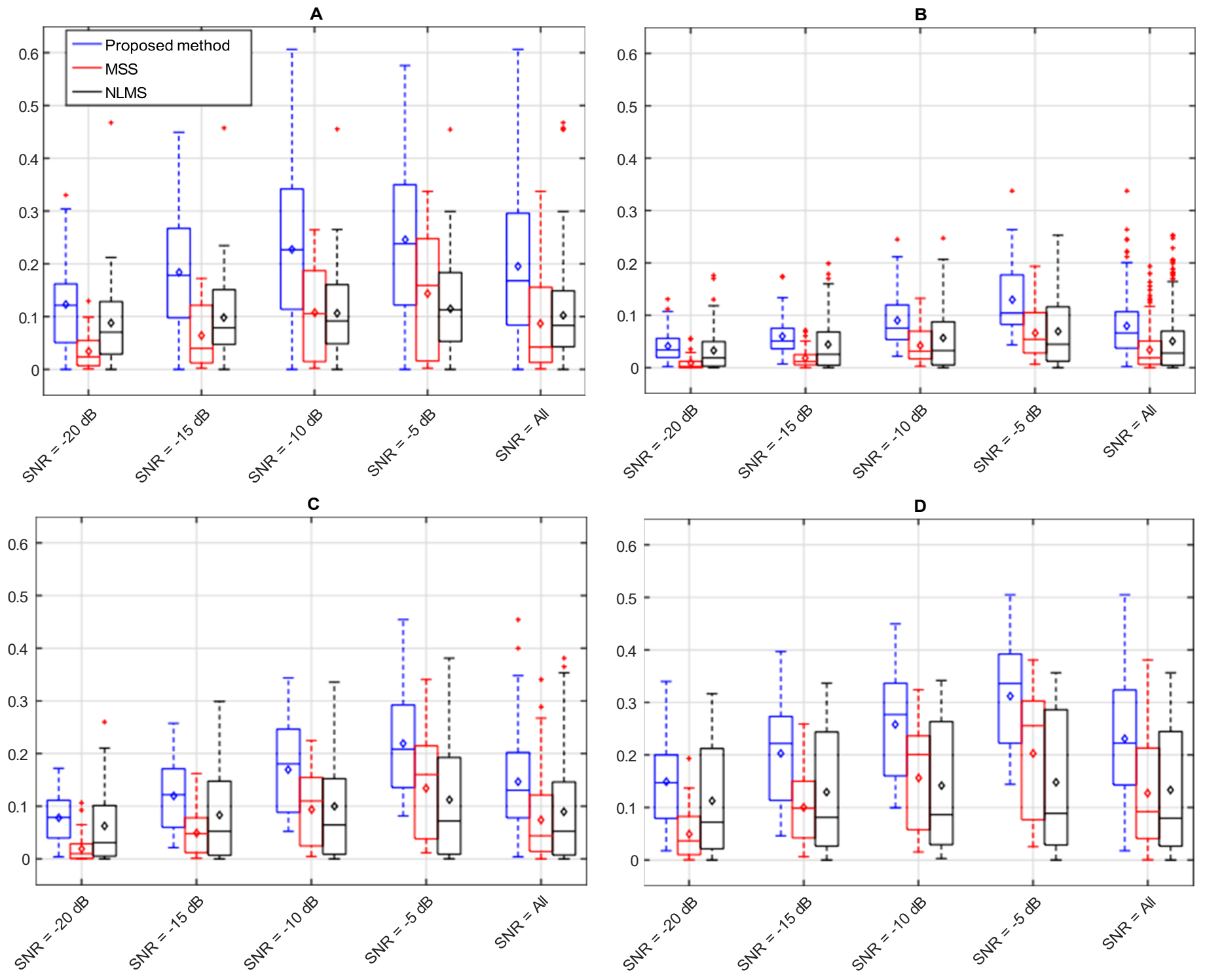

Figure 16: NCM (A), CSII$_{low}$ (B), CSII$_{med}$ (C) and CSII$_{high}$ (D) results considering all the noises (Ambulance Siren, Baby Crying, Babble and Street) for each SNR from database $D_F$. Each box represents the denoising performance results for each method evaluated. The color of the legend, associated to each evaluated method shown in the subfigure 16.A, refers to the same methods for all subfigures.

| **Algorithms** | **Multiplications** | **Additions** |
|---|---|---|
| Proposed method | $i_o M[FT(6 + 5K_N + 3K_S)$<br>$+K_N(3T + 2F) + K_S(T + F)]$<br>$+2TN\log_2(N)$ | $i_o M[FT(3K_N + 3K_S - 2)$<br>$-T(2K_N + 2K_S) - FK_S]$<br>$+4TN\log_2(N)$ |
| MSS | $13FT + 2TN\log_2(N)$ | $10FT + 4TN\log_2(N)$ |
| NLMS | $(3L + 1)f_s$ | $3Lf_s$ |

Table 2: Assessment of computational complexity in terms of the number of operations (multiplication and addition). Note that the number of operations has been calculated from the set of parameters on which the computational cost of each evaluated algorithm depends.

## 4. Conclusions and Future Work

In this work, we propose an incremental algorithm based on multichannel non-negative matrix partial co-factorization (NMPCF) for ambient denoising in auscultation focusing on high noisy environments with a Signal-to-Noise Ratio (SNR) lower than 0 dB. The first contribution applies NMPCF from a multichannel point of view assuming that ambient noise can be modelled as repetitive sound events found in two single-channel audio inputs simultaneously captured by means of different recording devices. The second contribution proposes an incremental algorithm, based on the previous multichannel NMPCF, that refines the estimated biomedical spectrogram through a set of incremental stages by eliminating most of the ambient noises that was not removed in the previous stage.

The optimization process, using a database that is not adapted to any specific patient's body or room, indicates that the best performance of the proposed method is obtained using a higher number of noise bases compared to the number of biomedical bases. It suggests that the energy distribution of the types of ambient noises analyzed is more complex compared to biomedical sounds due to the greater spectral diversity shown by the time-frequency structures of such noises.

Two databases have been created in order to simulate two different acoustic scenarios depending on whether the effects of propagation on the patient's body material and the acoustics of the room are inactive (ideal scenario) or active (real scenario). Each database is composed of a broad set of biomedical sounds (heart and lung) and ambient sounds mixed together with various levels of SNR.

The results obtained in the ideal scenario indicate that the proposed method outperforms MSS and NLMS, in terms of SDR, in most of the levels SNR showing the most robust SDR performance mainly in high noise environments (SNR∈[-20 dB,-10 dB]). Although NLMS is slightly better, in terms of SIR, compared to the proposed method, this fact occurs at the expense of losing a large amount of biomedical content. SDR and SIR improvements provided by MSS are ranked in the last position since the lowest levels SNR cause significant distortion of biomedical sounds at low spectral ranges. A remarkable advantage of the proposed method, compared to MSS and NLMS, is the high robustness of the acoustic quality of the estimated biomedical sounds when the two single-channel inputs suffer from a delay between them. As a consequence, the proposed method can be considered a more reliable ambient denoising approach since results obtained assuming ambient noises as repetitive sound events found in both the internal and external spectrograms report less dependence on the ratio of ambient noise to biomedical content or any temporal misalignment between the internal and external signal as occurs with MSS and NLMS.

From the real scenario, results show that the proposed method significantly outperforms MSS and NLMS for each SNR evaluated. The high and stable trend of SDR and SIR displayed by the proposed method for each SNR in com-parison with the rest of the methods is remarkable. Although NLMS shows better denoising performance compared to MSS, none of them can be considered competitive methods since their SDR results $\leq$ 1 dB when all ambient noises are averaged so, it implies a huge loss of biomedical content despite the fact that NLMS gets a great isolated biomedical signal. Comparing the denoising performance between the ideal and real scenario, it can be observed a high robustness of the proposed method compared to MSS and NLMS. Specifically, the average reduction of the SDR and SIR improvement suffered by the proposed averaging all ambient noises and taking into account the effects of the propagation of the patient's body material and the acoustics of the room is $\leq$ 1 dB and $\leq$ 5 dB while the same SDR and SIR reduction increases to 14 dB and 15 dB for MSS, and 18 dB and 21 dB for NLMS. Moreover, another set of metrics (NCM, $CSII_{low}$, $CSII_{med}$ and $CSII_{high}$) confirm that the proposed method provides the best ambient denoising performance by means of maximizing the ambient denoising at the expense of minimizing the loss of biomedical energy. Again, these results demonstrate that the multichannel co-factorization approach achieves a more adequate behavior in the removal of ambient noise compared to the previous spectral subtraction or adaptive filtering approaches. As a drawback, highlight the high computational cost of the proposed method mainly to the number of operation applied to the multiplicative update rules of the co-factorization process.

Future work will focus on two directions: (i) novel algorithms applied to the removal of some of the most disturbing acoustically active ambient noises in clinical emergency situations, such as noise inside a helicopter or ambulance when urgent monitoring is required, and, (ii) real-time multichannel non-negative matrix partial co-factorization approaches for biomedical ambient denoising.

## Funding

This work was supported by the Programa Operativo FEDER Andalucia 2014-2020 under project with reference 1257914 and the Ministry of Economy,

Knowledge and University, Junta de Andalucia under Project P18-RT-1994.

## Acknowledgment

The authors would like to thank Dr. Dinko Oletic and Dr. Vedran Bilas for sharing their lung recordings. The authors would like to thank the pulmonologist Gerardo Perez Chica from the University Hospital of Jaen (Spain) for his assistance related to ambient noises. Finally, we would like to thank the anonymous reviewers for their helpful and constructive comments that greatly contributed to improve the final version of the paper.